%% file: MBonavita_SPOTS_II_final.tex
\documentclass[article]{aa}  %[article]

\usepackage{longtable,lscape}
\usepackage{multicol}
\usepackage{natbib}
\usepackage{txfonts}
\usepackage{graphicx}
\usepackage{tabularx}
\usepackage[labelformat=simple]{caption,subcaption}
\usepackage{multirow}
\renewcommand\thesubfigure{\Alph{subfigure}}

\def\m2s2{\,m$^{2}$\,s$^{-2}$} %m2.s -2
\defcitealias{thalmann14}{Paper I}

\begin{document}

   \title{SPOTS: The Search for Planets Orbiting Two Stars}
    \subtitle{II. First constraints on the frequency of sub-stellar companions on wide circumbinary orbits}
    
\titlerunning{Constraints on the frequency of circumbinary planets in wide orbits}

   \author{M. Bonavita \inst{1,2}
           \and  S. Desidera \inst{2}
           \and  C. Thalmann \inst{3}
           \and  M. Janson \inst{4}
           \and  A. Vigan \inst{5}
           \and  G. Chauvin \inst{6}
           \and  J. Lannier \inst{6}
}

   \authorrunning{M. Bonavita et al.}

\institute{Institute for Astronomy, The University of Edinburgh, Royal Observatory, Blackford Hill, Edinburgh, EH9 3HJ, U.K.
   	\and INAF -- Osservatorio Astronomico di Padova, Vicolo dell'Osservatorio 5, I-35122, Padova, Italy
	\and Institute for Astronomy, ETH Zurich, Wolfgang-Pauli Strasse 27, 8093 Zurich, Switzerland
	\and Department of Astronomy, Stockholm University, AlbaNova University Center, 106 91 Stockholm, Sweden
	\and Aix Marseille Universit\'e, CNRS, LAM (Laboratoire d'Astrophysique de Marseille) UMR 7326, 13388, Marseille, France
	\and Institut de Plan\'etologie et d'Astrophysique de Grenoble, UJF, CNRS, 414 rue de la piscine, 38400 Saint Martin d'H\`eres, France}

\date{Received  / Accepted }

\abstract{A large number of direct imaging surveys for exoplanets have been performed in recent years, yielding the first directly imaged planets and providing constraints on the prevalence and distribution of wide planetary systems. However, like most of the radial velocity ones, these surveys generally focus on single stars, hence binaries and higher-order multiples have not been studied to the same level of scrutiny. This motivated the SPOTS (Search for Planets Orbiting Two Stars) survey, which is an ongoing direct imaging study of a large sample of close binaries, started  with VLT/NACO and now continuing with VLT/SPHERE. To complement this survey, we have identified the close binary targets in 24 published direct imaging surveys. 
Here we present our statistical analysis of this combined body of data.  
We analysed a sample of 117 tight binary systems, using a combined Monte Carlo and Bayesian approach to derive the expected values of the frequency of companions, for different values of the companion's semi-major axis. 
Our analysis suggest that the frequency of sub-stellar companions in wide orbit is moderately low ($\lesssim 13~\%$ with a best value of 6\% at 95\% confidence level) and not significantly different between single stars and tight binaries. 
One implication of this result is that the very high frequency of circumbinary planets in wide orbits around post-common envelope binaries, implied by eclipse timing (up to 90\% according to Zorotovic \& Schreiber 2013), can not be uniquely due to planets formed before the common-envelope phase (first generation planets), supporting instead the second generation planet formation or a  non-Keplerian origin of the timing variations.}


   \maketitle

\section{Introduction}

In the past decade, an increasing amount of effort has been spent on 
studying the formation and evolution of planets in the environment of
binary host star systems \citep[see e.g.\ the book 
``Planets in Binaries'',][]{2010ASSL..366.....H}.  
More than one hundred planets have been found in binary systems to date\footnote{exoplanets.org database \citep{2011PASP..123..412W}, www.exoplanets.eu \citep{2011AA...532A..79S}}.  Most
of these discoveries have been made with indirect detection methods
such as Doppler spectroscopy or transit photometry methods, which are
heavily biased towards planets with short orbital periods and, 
therefore, favour circumstellar (`s-type') configurations around 
individual components of wide binary systems.  
Despite this bias, about 20 of these planets have been found in circumbinary 
(`p-type') orbits encompassing tight binary systems, hinting at the 
existence of an extensive unseen population of 
circumbinary planets.

Direct imaging, on the other hand, is a powerful planet detection
technique particularly well suited to planets on wide orbits, which 
complements the limited parameter space of the indirect detection 
methods.  A number of direct imaging surveys have been published to
date \citep[e.g.,][]{2007ApJ...670.1367L,2012AA...544A...9V,
2013ApJ...773...73J, 2013ApJ...777..160B,2013AA...553A..60R, 2015ApJ...799..155D}, which
have resulted in the discovery of several planets 
\citep[e.g.][]{2010Natur.468.1080M,2010Sci...329...57L,
2013ApJ...774...11K,2013ApJ...763L..32C,2013ApJ...772L..15R} and 
brown-dwarf companions \citep[e.g.][]{2009ApJ...707L.123T,
2010ApJ...720L..82B, 2014ApJ...791L..40B}.  Such surveys typically reject
binary systems from their target sample. Although many previously unknown tight systems were still included in their target lists, the population of wide-orbit planets in such systems still remains largely unexplored.

To address this, the SPOTS project (Search for Planets Orbiting
Two Stars; \citepalias[\citealt{thalmann14}, hereafter][]{thalmann14} is conducting the first 
dedicated direct imaging survey for circumbinary planets.  Our 
long-term goal is to observe a large sample of young nearby tight binary systems with
the VLT NaCo, VLT SPHERE, and LBT/LMIRCAM facilities.  The NaCo-based first stage
of the survey, which comprises 27 targets, completed its exploratory 
observations in 2013 (Paper I)
and the follow-up observations to confirm the physical association
of planet candidates is in progress. Additional close binary  targets 
are being observed with the newly installed direct imaging instrument SPHERE \citep{2010ASPC..430..231B}
and with LMIRCAM at LBT in the context of the LEECH project \citep{2014IAUS..299...70S}, increasing
the sensitivity to planetary companions at close separation.
Although the survey is not yet completed, it has already yielded a first discovery: the sharp highly asymmetric features in the circumbinary protoplanetary disk around Ak~Sco imaged with SPHERE \citep{2016ApJ...816L...1J}.

A discussion of the survey's scientific 
background, observational strategy, and first results is presented in
Paper I. The 
scientific justification can be summarised in the following four main points:
\begin{itemize}
\item 
Theoretical and observational evidence suggests that circumbinary
planets constitute a significant fraction of the overall planet
population, and therefore merit exploration. 
\item 
With appropriate
target selection, the host binarity has no detrimental effects on 
observation and data reduction.  The detectability of planets around a
tight binary may in fact be superior to that around a single star of 
equal system brightness, since the greater total system mass is expected
to correlate with a greater amount of planet-forming material. 
\item
Dynamic interactions with the host binary can launch circumbinary 
planets that formed or migrated close to the system centre onto wide
orbits, where they are more easily imaged.
\item
Measuring differences in
the planet demographics between circumbinary and single-star target
samples may bring new insights into the physics of planet formation and
evolution that would be inaccessible to surveys of single stars only.
\end{itemize}
Details and references for these claims are listed in 
\cite{thalmann14}.

Here, we present a statistical analysis of the combined body of 
existing high-contrast imaging constraints on circumbinary planets to 
complement our ongoing survey.  Indeed, while several of the available surveys
intended to avoid binaries, or at least close visual binaries, the census of stellar
multiplicity was highly incomplete at the time of the execution of the observations.
The direct imaging surveys provided themselves the best census of close visual binaries,
with each survey contributing typically with several new discoveries.

For this purpose, we searched the target
lists of 23 published direct imaging surveys, looking for tight binaries,
collected their contrast curves, and compared them to synthetic 
circumbinary planet populations using the \texttt{QMESS} code 
\citep{2013PASP..125..849B}.  The target sample is presented in 
Section~\ref{sec:sample}, the stellar and binary properties
in Section~\ref{sec:prop} and the statistical analysis
is described in Section~\ref{sec:main_stat}. Finally the results are summarised and discussed in Section~\ref{sec:discussion}.

\section{Target samples}
\label{sec:sample}
\subsection{The circumbinary sample}
Our initial sample was built merging the target lists of the several recent deep imaging surveys with sensitivity adequate for detection of giant planets.
Among these are some of the largest deep imaging surveys performed to date, such as the VLT/NaCo large program by \cite{2015AA...573A.127C} (NLP), the PALMS (Planets around Low-mass Stars) survey \citep{2015ApJS..216....7B}, the SEEDS (Strategic Exploration of Exoplanets and Disk with Subaru) survey \citep[][B13 and J13, respectively]{2014ApJ...786....1B, 2013ApJ...773...73J} and the Gemini NICI Planet-Finding Campaign \citep[][N13 and BN13, respectively]{2013ApJ...776....4N,2013ApJ...777..160B}. 
The main characteristics of all the surveys considered in this paper are reported in Table \ref{tab:survey}.
To these, we added also the low-mass spectroscopic binary CHXR~74, which orbit has been constrained by \cite{2012AA...537A..13J} (JJ12).

We also included some target from a HST/NICMOS survey of 116 young ($<~30~Myrs$) nearby ($<~60~pc$) stars \citep[Song et al. private communication, see also][]{2006ApJ...652..724S}. 
Each target was observed at two spacecraft roll angles in successive HST orbits.
After standard cosmetics correction, the two roll angle images were recentered and subtracted to suppress the stellar Light contribution. Additional Fourier filtering was applied to remove PSF low-spatial frequencies to search for faint point-like sources in the star vicinity.
Detection limits and maps were derived using a 5x5 pixels sliding box over the whole image and flux calibrated considering the standard NICMOS photometric calibration in the F160W observing filters (please refer to: http://www.stsci.edu/hst/nicmos/performance/photometry)

\begin{table*}
   \caption[]{Characteristics of the surveys considered to build the circumbinary (CBIN) sample.}
     \label{tab:survey}
       \centering
       \begin{tabular}{lcccccl}
         \hline
         \noalign{\smallskip}
         Source  &  Instrument & Technique$^1$ & Filter & $N_{Srv}^2$ & $N_{CBIN}^3$   & Reference \\

         \noalign{\smallskip}
         \hline
         \noalign{\smallskip}
%% M05        & VLT/NACO	    & SatDI	    & K/H           & 28  & \cite{2005ApJ...625.1004M}    \\ % Nearby Young Stars
 L05        & HST/NICMOS    & COR	    & ~H(1.4-1.8)	& 45  &  6   & \cite{2005AJ....130.1845L} \\ % Nearby Young Stars
 B06        & VLT/NACO      & COR       & $K_S$/H       & 17  &  3    & \cite{2006ApJ...652.1572B} \\ % Nearby Young Stars
 B07        & VLT-NACO/MMT  & SDI	    & H 	        & 45  &  7    & \cite {2007ApJS..173..143B} \\ % Nearby Young Stars
 K07        & VLT/NACO	    & DI	    & L 		    & 22  &  4    & \cite{2007AA...472..321K} \\ % Members of associations
 GDPS       & GEMINI/NIRI   & SDI	    & H 		    & 85  &  8    & \cite{2007ApJ...670.1367L} \\ % Nearby Young Stars
 CH10       & VLT/NACO      & COR	    & H/$K_S$		& 91  &  9    & \cite{2010AA...509A..52C} \\ % Nearby Young Stars
 H10        & Clio/MMT      & ADI       & L'/M          & 54  &  3   & \cite{2010ApJ...714.1551H}\\
 JB11       & GEMINI/NIRI   & ADI       & K/H           & 18  &  3    & \cite{2011ApJ...736...89J} \\ % Early type stars
% D12        & VLT/NACO      & DI/ADI    & L             & 16  &  1    & \cite{2012AA...539A..72D} \\ % M dwarfs
 JJ12       & VLT/NACO      & DI        & $K_S$         & 1   &  1    & \cite{2012AA...537A..13J} \\
 V12        & VLT/NACO, NIRI& ADI       & $K_S$/H'/CH4  & 42  &  3  & \cite{2012AA...544A...9V} \\ % Early type stars
 R13        & VLT/NACO      & ADI       & L'            & 59  &  3    & \cite{2013AA...553A..60R} \\ % Mostly early type stars
 B13        & SUBARU/HiCiao & DI/ADI/PDI& H             & 63  &  6    & \cite{2014ApJ...786....1B} \\ % Members of associations
 J13        & SUBARU/HiCiao & ADI       & H             & 50  &  4    & \cite{2013ApJ...773...73J} \\ % Dbris disks stars
 Y13        & SUBARU/HiCiao & ADI       & H/$K_S$       & 20  &  3    & \cite{2013PASJ...65...90Y} \\ % Pleiades
 N13        & GEMINI/NICI   & ADI/ASDI  & H             & 70  &  4    & \cite{2013ApJ...776....4N} \\ % NICI Early type stars
 BN13       & GEMINI/NICI   & ADI/ASDI  & H             & 80  &  4    & \cite{2013ApJ...777..160B} \\ % NICI Young Moving groups
 JL13       & GEMINI/NICI   & DI/ADI    & $K_S$         & 138 &  5    & \cite{2013ApJ...773..170J} \\ % Sco-Cen
 L14        & GEMINI/NIRI   & DI/ADI    & $K_S$         &  91 & 18    & \cite{2014ApJ...785...47L}\\ % upper scorpio
 SONG       & HST           & ADI       & H             & 116 & 14    & Song et al.\ priv.\ comm. \\ % Song
 M14        & VLT/NACO      & ASDI      & H             &  16 &  1    & \cite{2014AA...566A.126M} \\ % NACO-ASDI
 NLP        & VLT/NACO      & DI/ADI    & H             & 110 &  8    & \cite{2015AA...573A.127C} \\ % Naco LP
 D15        & GEMINI/NIRI   & DI        & $K_S$         &  64 &  4    & \cite{2015ApJ...799..155D} \\ % Taurus
 B15        & SUBARU/HiCiAO & DI/ADI    & $K_S$         &  31 &  5    & \cite{2015ApJS..216....7B} \\ % PALMS
            & KECK/NIRC2/N  & DI/ADI    & H             &  59 &  3    &     \\
 L15        & VLT/NACO      & ADI       & L'            &  58 &  10    & Lannier et al. 2016 (submitted)  \\ %Lannier in 
 
 \noalign{\smallskip}
         \hline
      \end{tabular} 
\caption*{\footnotesize $^1$Techniques: {\bf COR} = Coronagraphy; {\bf SDI} = Spectral Differential Imaging; {\bf DI} = Direct Imaging; {\bf ADI} = Angular Differential Imaging; {\bf PDI} = Polarized Differential Imaging; {\bf ASDI} = Angular and Spectral Differential Imaging.\\
$^2$Total number of targets included in the original survey;  $^3$Number of stars considered in our study.}
\end{table*}

For all the targets an extensive search for multiplicity was performed in binary catalogues 
such as the Hipparcos and Tycho Catalogues \citep{1997ESASP1200.....P}, 
the Catalogue of the Components of Double and Multiple Stars  \citep[CCDM][]{2002yCat.1274....0D},
the Washington Visual Double Star Catalogue \citep[WDS][]{1997AAS..125..523W}, 
the 9th catalogue of spectroscopic orbits \citep[SB9][]{2004AA...424..727P},
the SACY database \citep{2006AA...460..695T}, the Geneva-Copenhagen survey \citep{2004AA...418..989N}.
We also considered the literature on individual targets as well as from the direct imaging surveys 
themselves, which resolved for the first time a number of pairs, making the input papers the 
best sources to be used to identify close visual binaries.
Ambiguous cases such as candidate binaries with astrometric accelerations only or with position
above sequence of coeval stars in colour-magnitude diagram are not included in our sample of binaries.
We also note that several of the targets of imaging surveys are lacking radial velocity monitoring, thus
the census of spectroscopic binaries is likely incomplete.

When searching for circumbinary planet hosts in such samples, one must take into account that most of these surveys includes severe selection biases against binary targets. 
Most surveys in fact excluded known binaries with separations smaller than 2 arcsecs.
Nevertheless, a significant number of binary and multiple targets are found in this surveys, not being known at the time of 
the target list compilation, or resolved for the first time during the searches themselves.

Of course, wide binaries are not suited to a search for circumbinary planets.
We fixed as a limit for our investigation the systems for which the  inner limit of dynamical stability
for circumbinary planets (see Sect.~\ref{sec:bin_par} for definition and determination) is smaller than 50 au.
This limit roughly corresponds to the expected truncation limit of the circumbinary disk.
The adopted limit is significantly larger than the dynamical stability limits for the circumbinary
systems discovered by Kepler but it can be considered as conservative when looking at the 
properties of some binaries hosting well-studied circumbinary disks such as GG Tau A \citep[$a \sim$ 60 au, ][]{2011AA...530A.126K} and SR24N \citep[$a \sim$ 32 au, ][]{2005ApJ...619L.175A}.

Therefore, while the adopted limit is somewhat arbitrary, it appears reasonable for the identification of a sample
of systems for which the presence of circumbinary planets is possible and worth to be explored.

With such selection criteria,  a total of 139 targets were selected. Taking into account the overlap between the various surveys considered, our final sample for the search for circumbinary planets (hereafter CBIN sample) includes 117 unique systems. 

The stellar and binary parameters of the stars in the CBIN sample are derived following the prescriptions described in Sec.~\ref{sec:prop} and are listed in Table~2.

It is interesting for the purposes of our statistical analysis and for comparison
with other results (e.g., from Kepler space mission) to obtain an ensemble view of the properties of the sample.
To this aim, Fig.~1 shows histograms and plots of several relevant
parameters, derived as described in Sec.~\ref{sec:bin_sel}.
As expected, the sample is dominated by young stars, with median age $\sim 50$~Myr. Nevertheless, several old stars are present, mostly tidally-locked binaries originally classified as young due to their high activity levels.
The median distance of the systems is 45~pc, with a significant number of objects (25\%) at distances larger than 100 pc, mostly members of Sco-Cen groups.
The total system mass lies between 0.22 to 20.8~$M_{\odot}$, with a median value of 1.34~$M_{\odot}$.
The distribution of critical semi-major axis has a median value of 10~au, with 48~\% of systems with $a_{crit}<$~10~au. Binaries at larger $a_\mathrm{crit}$ are under-represented in the sample with respect to unbiased samples due to the exclusion of previously known close visual binaries in most of the imaging surveys.
The mass ratio distribution is fairly uniform, with a median value of 0.61.

\begin{figure*}[h!]
\centering
\label{fig:sample}
\includegraphics[width=90mm]{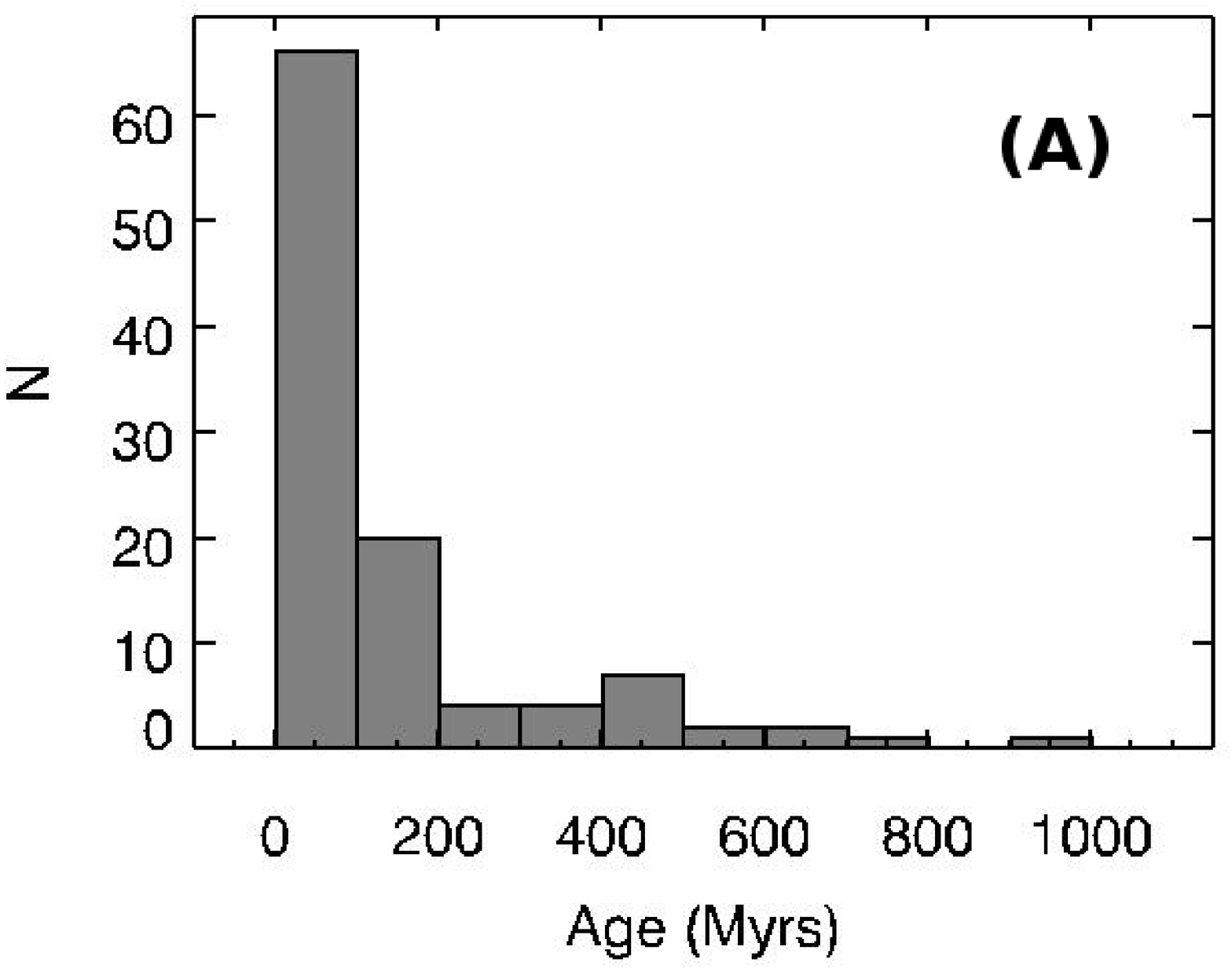}\hspace{1em}\label{fig:age}
\includegraphics[width=90mm]{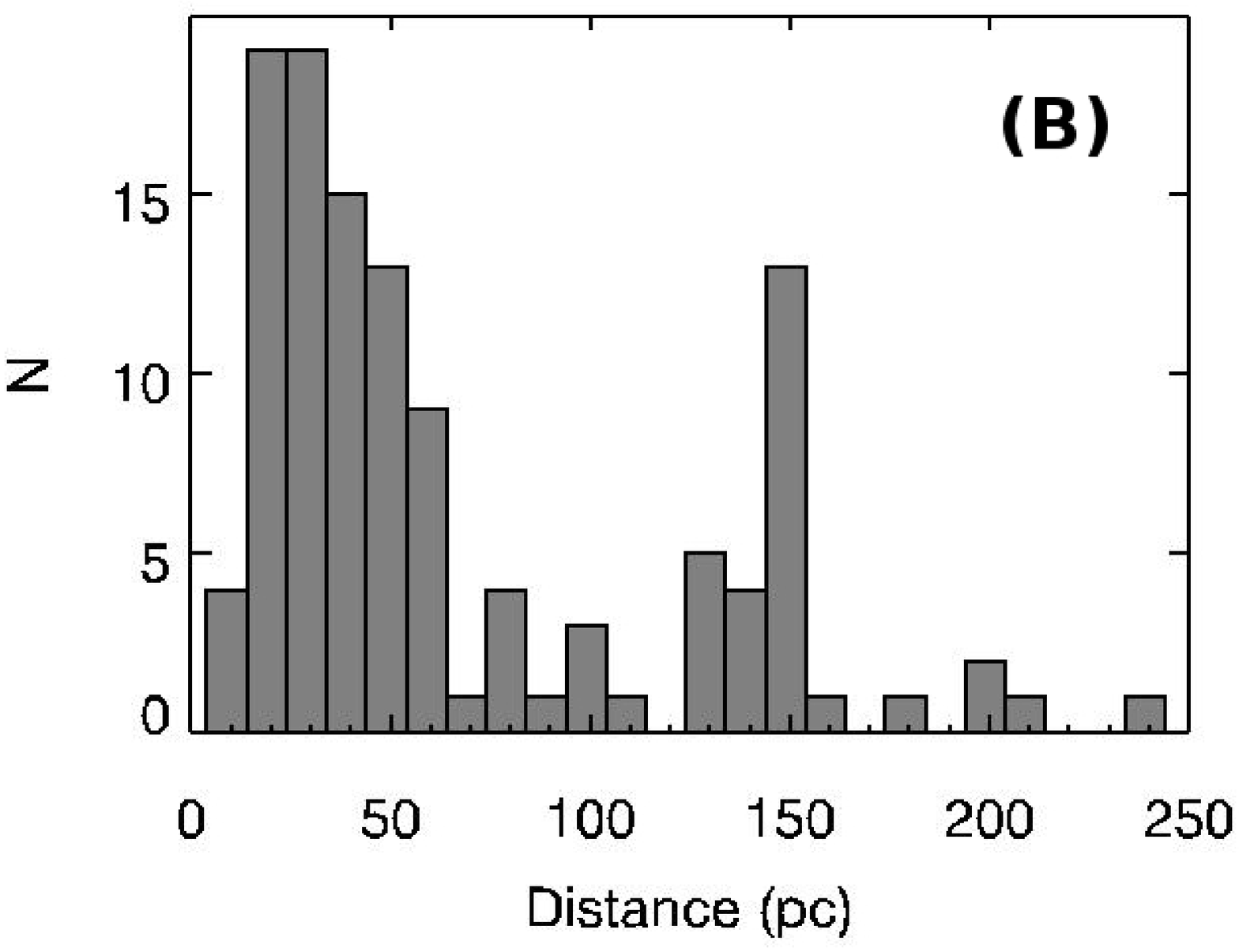}\label{fig:dist}
\includegraphics[width=90mm]{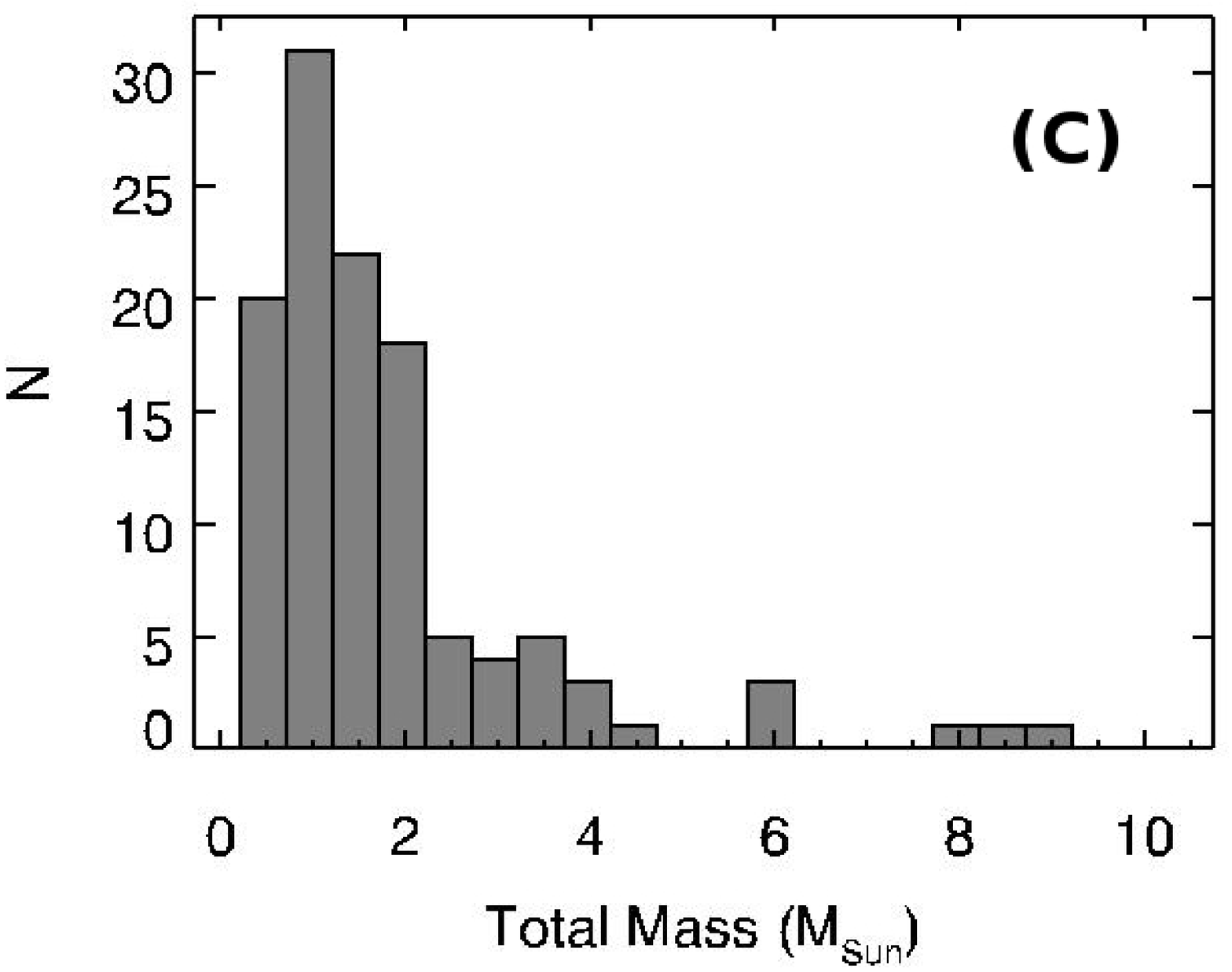}\hspace{1em}\label{fig:acrtitmratio}
\includegraphics[width=90mm]{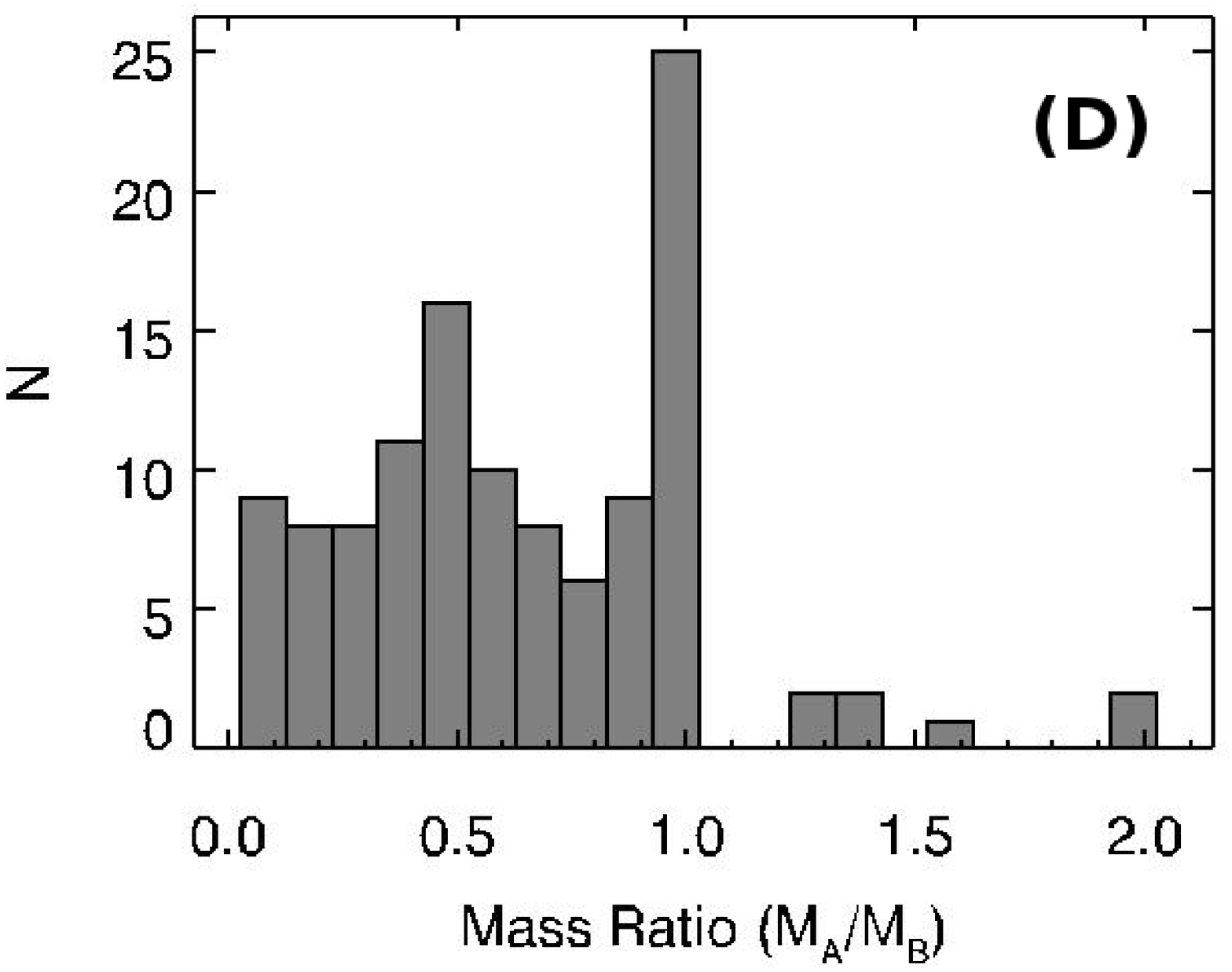}\label{fig:mtot}
\includegraphics[width=90mm]{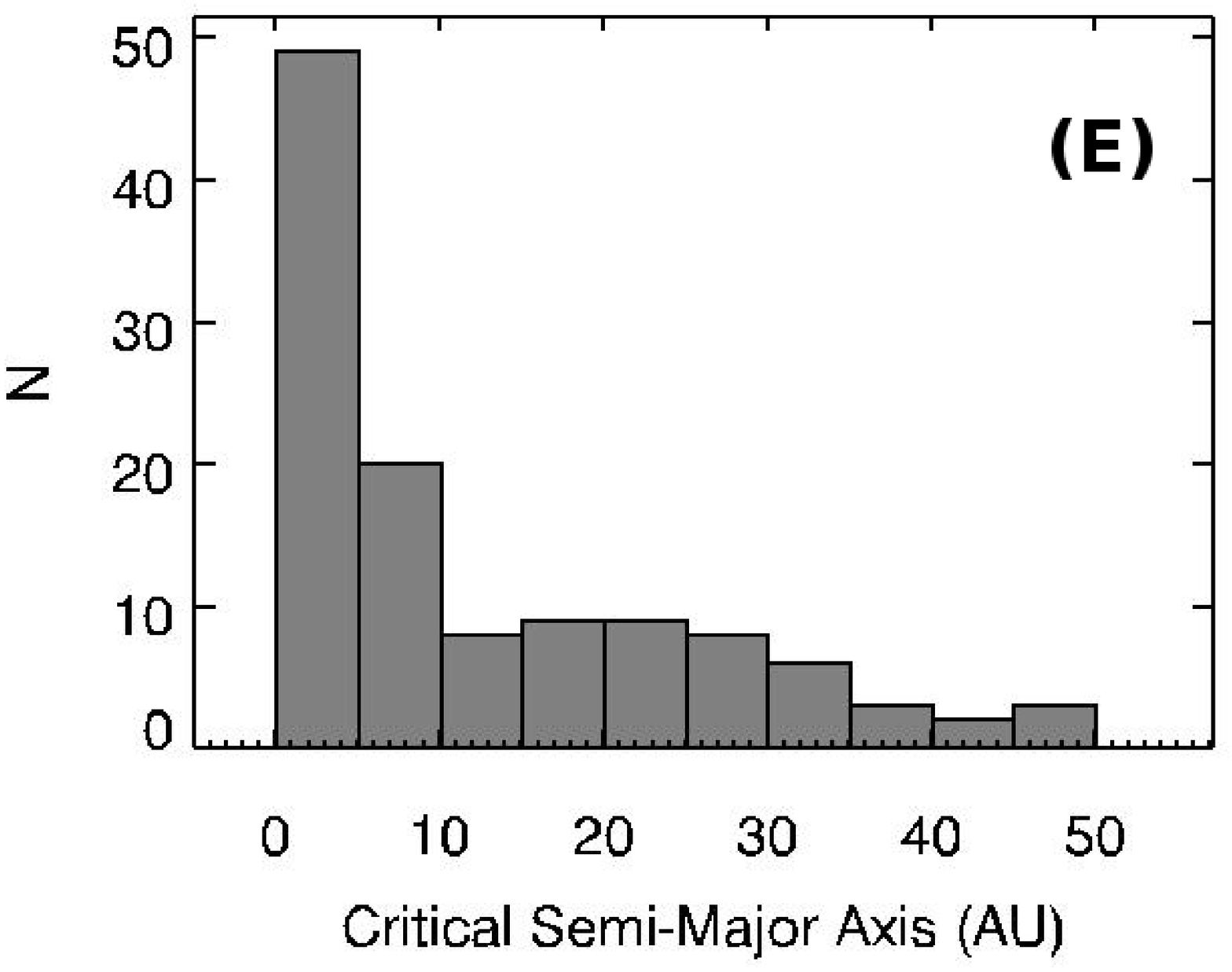}\hspace{1em}\label{fig:acrit}
\includegraphics[width=90mm]{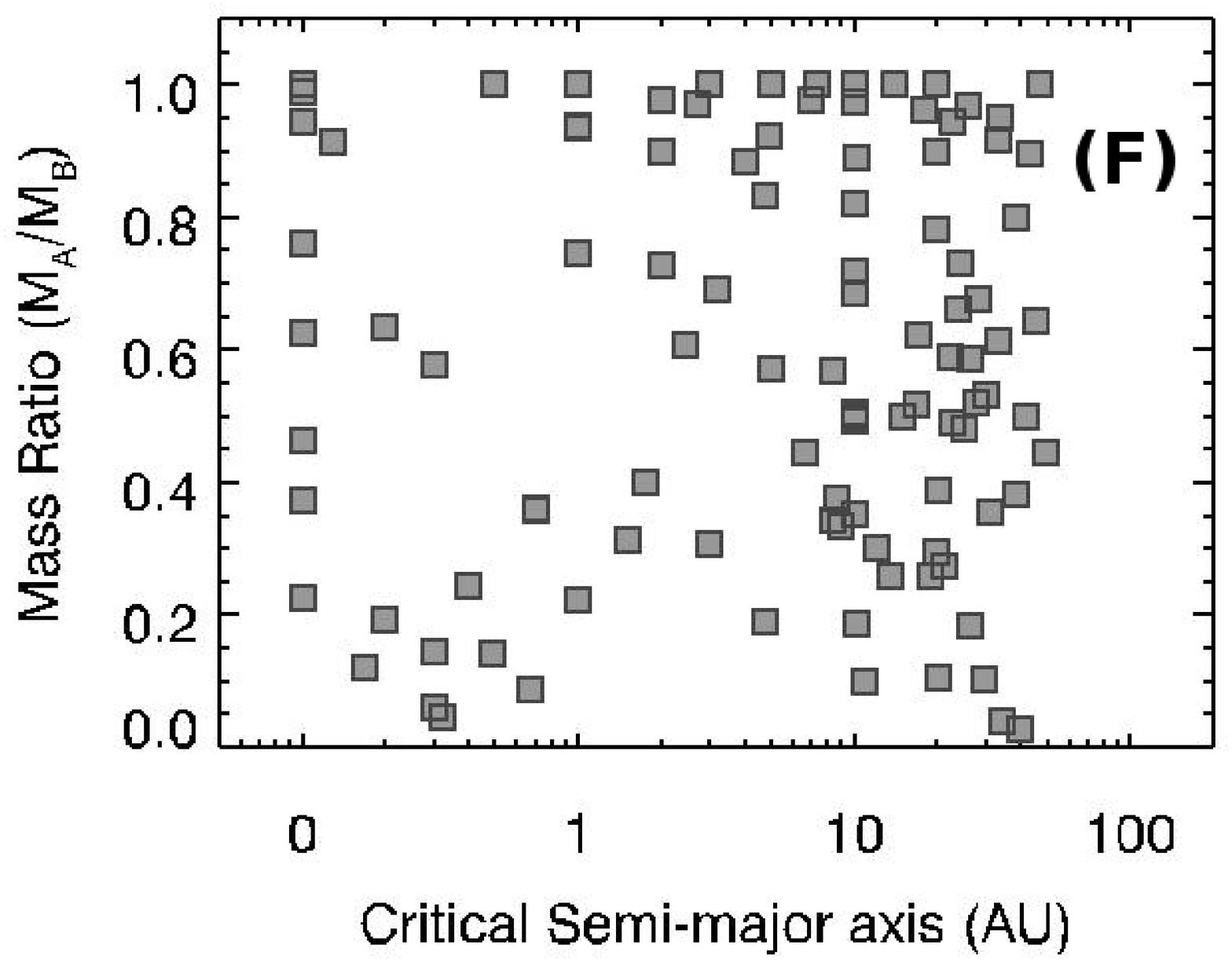}\label{fig:acrtitmratio}
\caption{Properties of the stars in the CBIN sample. {\bf [A]}: histogram of stellar ages; {\bf [B]}: histogram of stellar distances; {\bf [C]} histogram of total system masses; {\bf [D]}: histogram of the system mass-ratio; {\bf [E]}: histogram ofcritical semi-major axis for dynamical stability of planetary companions ($a_\mathrm{crit}$). {\bf [F]}: Inner limit for circumbinary planet stability ($a_\mathrm{crit}$) vs binary mass ratio.}
\end{figure*}

\subsection{The control sample}
\label{sec:ss_sample}

In order to ensure a consistent comparison of our results with those obtained for single stars, we carried an independent analysis of the sample described by \cite{2014ApJ...794..159B}.
All the binaries used for our analysis were removed from the sample, together with those targets for which the detection limits were not available.
We also removed from the comparison sample the stars with stellar companions within 100~au.
As suggested by \cite{2007AA...468..721B} and, more recently by \cite{2010ApJ...709L.114D}, systems with separation $>~100$~au  are in fact indistinguishable from single stars as far as the initial conditions and end product of planet formation are concerned. 
With these assumptions, the final control sample (hereafter SS sample) includes 205 stars.

\section{CBIN sample properties}
\label{sec:prop}

The CBIN sample is quite heterogeneous in terms of stellar and binary properties, as expected considering the original selection criteria in the parent surveys, which are focused in some cases of specific types of stars (low mass stars, early type stars, specific young moving groups), the presence or not of biases against specific types of binaries, etc.
In this section, we present our determination of stellar and binary parameters 
for the systems included in our sample.

\subsection{Stellar parameters}
\label{sec:bin_sel}

\subsubsection{Stellar Ages}

Even if their evolution is not completely understood  
\citep[see][]{2008ApJ...683.1104F},  giant planets are 
in fact thought to be more luminous at young ages, their luminosity 
fading with time, as they cool down 
\citep[see][]{2003AA...402..701B, 2007ApJ...655..541M}. 
Thus, observing  younger targets increases the probability to find smaller 
companions by raising the planet/mass contrast, especially in the IR domain. 
Therefore, most of the original target lists for the surveys we considered were assembled on the basis of the young ages. 

In the past few years, significant efforts were devoted to the identification 
of nearby young stars and to the determination of their
basic parameters.
However, the determination of stellar age is still a challenging task \citep{2014prpl.conf..219S} 
and stellar multiplicity represent an additional source of complications due to blending of the spectral features
and lack of spatially resolved fluxes for most of the systems studied in the present paper.
Furthermore, in very close binaries the components are tidally locked and so they have a short rotation period, thus mimicking some of the characteristics of young stars, such as high levels of chromospheric and coronal activity.
There are also claims that Lithium abundance, another widely used
age indicator, is altered in tidally-locked binaries \citep{1992AA...253..185P}.

There are several cases of stars included in the direct imaging surveys being classified as young
thanks to their high level or chromospheric and coronal activity 
but the subsequent identification of their nature as 
close spectroscopic binaries suggest that these are due to
tidal locking and not to young age. In these cases, the determination of the
stellar ages is very critical, especially when the lack of orbital solution
prevents the study of the system kinematic. In some cases, we conservatively adopt 
an age of 4 Gyr, given the lack of specific constraints on stellar age.
In some other cases, multiplicity was not known or in any case not taken into account in the
derivation of stellar properties, resulting in biased parameters (e.g., photometric distances
and then kinematic parameters).

In general, we followed the procedures described in \citet{2015AA...573A.126D} to derive stellar ages.
For field stars, stellar ages were obtained from a variety of age indicators
(lithium, chromospheric emission, coronal emission, rotation period,
kinematic, isochrone fitting), exploiting measurements and age calibrations
published after the original papers presenting the direct
imaging surveys. For this reason, in several cases the system ages adopted in this work differ
from those of the original papers. 
For close binary systems evolved through mass exchanges phase, ages
and individual masses were taken from papers dedicated to the study
of these objects. 

Age is easier to determine in young associations, because a variety of 
stellar dating techniques can be used for stars of different masses 
(stellar models for low-mass stars and massive evolved stars, lithium, etc,) 
or for the association as a whole (kinematic age derived from
relative velocities and position of the members).
 
The membership of the targets to various young associations and clusters was taken from several
literature sources \citep{2004ARAA..42..685Z,2008arXiv0808.3362T,2011ApJ...732...61Z,2013ApJ...762...88M} 
and on studies of individual objects.
Following the most recent results published in the literature in the last year, the
ages of several young moving groups were revised with respect to those adopted in
\citet{2015AA...573A.126D} and in Paper I.
For $\beta$ Pic, Tuc-Hor, Columba, AB Dor, TW Hya associations and $\eta$ Cha open cluster
we adopt the ages from \citet{2015MNRAS.454..593B}.
For Argus-IC 2391, we adopt the Li-depletion boundary age by \citet{2004ApJ...614..386B}, considering the ambiguities
in the isochrone fitting discussed in \citet{2015MNRAS.454..593B}.
For Sco-Cen groups, we adopt the ages from \citet{2012ApJ...746..154P}, as already done in \citet{2015AA...573A.126D}. They are based on the same
technique employed in  \citet{2015MNRAS.454..593B}, even if there are differences in some details of the isochrone fitting procedure. 
The resulting age ranking is also consistent with the result that the Lower Crux Centaurus group (LCC) is younger than $\beta$ Pic moving group (MG) members, as found by \citet{2012AJ....144....8S} from Li EW.
To be consistent with the upward revision of ages of most moving groups, we also revise the age of the Carina-Near moving group to 250 Myr. This is consistent with the recent gyro-chronology age of the nearly coeval Her-Lyr association \citet{2013AA...556A..53E}, although we do not have targets from this last group in our list. 
For Pleiades and Hyades open clusters we adopt 125 and 625 Myr, respectively. For Castor and Ursa Major
moving groups we adopt 320 and 500 Myr, respectively.

Details of the age indicators and membership to groups for individual targets are provided
in Appendix~\ref{app:notes},
The ages of moving groups as described above were also adopted for the members included in the comparison sample of single stars considered in the statistical analysis in Sec.~\ref{sec:cmp_freq}

\subsubsection{Stellar Distances}

Trigonometric distance from Hipparcos New Reduction 
\citep{2007AA...474..653V} or other individual sources were adopted when available.
For other members of groups  \cite{2008arXiv0808.3362T} photometric+kinematic
distances were adopted.
For members of Upper-Scorpius without trigonometric parallax, a distance of 145 pc is adopted.
For field stars without trigonometric parallax, photometric distances were
derived using empirical sequences for different ages determined from
members of moving groups, as described in \citet{2015AA...573A.126D}.

\subsubsection{Stellar masses} 
\label{sec:masses}

Stellar masses were derived in most cases through stellar models for the adopted ages.
In some case individual dynamical masses or mass ratio are available
from orbital solution and we took into account this information.
For the spectroscopic binaries for which only minimum mass of the companion is 
available from the orbital solution, we adopt this value to derive the
critical semi-major axis for dynamical stability see Sect.~\ref{sec:bin_par}).
For the spectroscopic binaries for which minimum mass is not available 
(e.g. only indication for short period RV variations without orbital 
solution), we adopt a mass equal to half of that of the primary for the computation of
the dynamical stability limit.
Fig.~1C and 1D show the histograms of the total mass ($M_A+M_B$) and of the mass ratio ($q=M_B/M_A$), respectively, for the systems in the CBIN sample.  Note that for the few systems where the secondary is a tight pair (see Sec.\ref{sec:triple} and Tab.~\ref{tab:widebin} for details) the total mass of the two components was considered, thus resulting in a value of $q>1$.

\subsection{Binary parameters}
\label{sec:bin_par}

The properties of the systems included in the CBIN sample are listed in Table~2.
References and details on individual systems are provided in Appendix~\ref{app:notes}.
When the complete orbital solution is known, semi-major axis and eccentricity are listed.
For systems for which no reliable semi major axis was available, we made the estimation that $a$(au) $\sim$ $\rho$(arcsec)$d$(pc). This relies on the assumption of a flat eccentricity distribution, based on the results of \cite{2010ApJS..190....1R}.

For spectroscopic binaries the masses as described in Sect. \ref{sec:masses} were adopted.

Tab.~2 also reports the values of the critical 
semi-major axis for dynamical stability ($a_\mathrm{crit}$), calculated following the approach of \cite{1999AJ....117..621H},
For the circumbinary case this inner limit for the stability is given by:
\begin{eqnarray}
\label{eqn:acrit_cb} 
a_{crit} = a_{CB}  = \left(1.60 + 4.12 ~\mu + 5.10~e_b\right) a_b  \nonumber \\
 + \left(-4.27 ~\mu ~e_b -  5.09 ~\mu^2\right) a_b \nonumber \\
  + \left(-2.22 ~e_b^2 + 4.61 e_b^2~\mu^2\right) a_b.
\end{eqnarray}

\noindent In the equation we assume $\mu=\frac{M_B}{M_A+M_B}$, where $M_A$ is the mass of the primary star, $M_B$ the mass of the secondary and  $a_{bin}$ and  $e_{bin}$ are  
the semi major axis and  the eccentricity of the binary orbit.
In agreement with the assumption used for the semi-major axis calculation, an eccentricity value of 0.5 was adopted for the systems for which no information on the orbit was available.  

We choose $a_\mathrm{crit}$ as a reference value because it is a physical quantity that better represents the 
dynamical effects due to a companion on planet formation and stability, including both the 
orbital parameters and mass ratio. Only planets outside the $a_\mathrm{crit}$ limit for circumbinary planets were
considered in the statistical analysis.

\subsubsection{Higher order systems}
\label{sec:triple}
There are several cases among our targets showing higher order multiplicity.
Five systems (Algol, TWA5, BS Ind, V815 Her and HIP 78977) are tight triple systems with an inner pair with period shorter than 5 days and an external component with semi-major axis smaller than 3~au.
In these cases, the direct imaging data would be able to detect planets around the three components.
The critical semi-major axis for circumbinary planets was derived in these cases considering the sum of the masses of the inner pair, the mass of the outer component and the outer orbital parameters. 

There are also several cases of hierarchical systems with an additional component at wide separation (Table \ref{tab:widebin}).
In these cases, we considered the dynamical effects on possible circumbinary planets considering the tight binary as a single star with a mass resulting from the sum of the individual components. 
The limit for the presence of circumbinary planets due the outer companion(s) is therefore derived using the equation by \cite{1999AJ....117..621H} for circumstellar planets:

\begin{eqnarray}
 \label{eqn:acrit_cs}
 a_{crit} = a_{CS} = \left(0.464 -0.38 ~\mu + 0.361~e_b\right) a_b  \nonumber \\
 + \left(0.586 ~\mu ~e_b +0.150 ~e_b^2\right) a_b \nonumber \\
 + \left(-0.198 ~\mu ~e_b^2\right) a_b.
\end{eqnarray}

For the 31 systems listed in Tab.~\ref{tab:widebin} this outer stability limit is smaller than the maximum value considered for the planetary semi-major axis (1000~au). Therefore for these targets both the inner and outer limit for the stability have been considered for the statistical analysis (Sect. \ref{sec:stat}). 

The few cases of compact triple systems for which the stability limit due to the presence of the outer component is smaller than the limit for circumbinary planets around the central pair were removed from the sample.

\begin{table*}
\setcounter{table}{2}
   \caption[]{Additional wide companion around the close pairs in the CBIN sample.\label{tab:widebin}}
       \centering
       \begin{tabular}{llccccccc}
         \hline
         \noalign{\smallskip}
\#$^1$  & Star ID     &   $M_{Target}^2$  &  $M_{Outer}^3$ & $\rho$ & a    & e & $a_{CS}^4$ & Notes \\
        &             &   $M_{\odot}$      &  $M_{\odot}$   & ('')   & (au) &   &  (au)      & \\
         \noalign{\smallskip}
         \hline
         \noalign{\smallskip}
6 & HIP 4967          &  0.88   & 1.22    & 25.6        &   765     &   --       &   81  &  \\ % NEW
10 & HIP 12413         &  2.63   & 0.40    & 23.8        &   947     &   --       &  159  &  \\ % NEW
12 & HIP 12638               &  1.19   & 0.80    & 14.57       &   662     &    --      &   87  &  \\ % updated
13 & HIP 13081               &  1.16   & 0.16    &  20.0       &   493     &    --      &   83  &  \\ % updated
23 & RX J0415.8+3100         &  1.16   & 0.62    &  0.95       &   190     &     --     &   26  &  \\ % updated
25 & HIP 21482               &  1.03   & 0.67    &   126       &  2268     &     --     &  300  &  \\ % updated
26 & GJ 3305                 &  1.35   & 1.60    &  66.0       &  1942     &    --      &  217  &  \\ % updated
29 & HIP 23296               &  1.79   & 0.09    &  9.17       &   455     &    --      &   82  &  \\ % updated 
30 & HIP 23418               &  0.41   & 0.25    &  1.37       &    34     &    --      &    4  &  \\ % updated
33 & AB Dor~AC               &  0.96   & 0.32    &  9.0        &   136     &    --      &   21  &  \\ % updated 
34 & AB Dor~Bab              &  0.32   & 0.96    &  9.0        &   136     &    --      &   11  &  \\ % updated  
38 & HIP 35564               &  2.19   & 2.40    &  9.0        &   285     &    --      &   32  & quintuple system        \\ % updated 
41 & GJ 278~C                &  1.20   & 4.83    &  72         &  1073     &    --      &   82  & Castor, sextuple system \\ % updated 
43 & HIP 39896~A               &  1.00   & 0.72    &  14         &   298     &    --      &   38  & close pair of M dwarfs  \\ % updated 
44 & HIP 39896~B              &  0.72   & 1.00    &  14         &   298     &    --      &   32  & quadruple \\ % updated 
52 & HIP 49669               &  3.70   & 1.10    &  175        &  4165     &    --      &  644  & quadruple \\ % updated 
59 & HD 102982               &  2.18   & 0.33    &  0.90       &    56     &    --      &    9  &  \\ % updated  % updated 
71 & HIP 72399               &  1.12   & 0.71    &  11.0       &   507     &    --      &   67  &  \\ % updated  % updated 
75 & HIP 76629               &  1.23   & 0.4     &  10.2       &   393     &    --      &   60  &  \\ % updated  % updated 
83 & 1RXS J160210.1-2241.28  &  1.35   & 0.53    &  0.300      &    43     &    --      &    6  &  \\ % V1154 Sco% updated 
90 & HIP 79097               &  3.06   & 0.75    &  0.814      &   163     &    --      &   26  &  \\ % inclusion TBC, m_A TBC % updated 
93 & HIP 79643~B              &  1.05   & 2.10    &  1.24       &   262     &    --      &   25  &  \\ % L14  % updated 
95 & HIP 84586               &  2.05   & 0.25    &  33         &  1038     &    --      &  178  &  \\ % updated 
97 & HIP 86346               &  1.23   & 0.30    &  19.6       &   590     &    --      &   94  &  \\ % updated 
99 & CD-64 1208~A            &  1.31   & 1.60    &  70         &  1998     &            &  222  &  \\  % updated 
102 & HIP  94863              &  1.46   & 0.26    &   9.4       &   394     &    --      &   65  &  \\ % updated 
104 & HIP 97255               & $\sim$1.40 & 0.60 &  9.90       &   307     &    --      &   44  &  \\ % updated  
105 & 2MASSJ19560294-3207186 &  0.30  & 0.55 &  26.0       &  1430     &    --      &  140     &  \\  %% NEW giusto 
110 & HIP 105441              &  1.27   & 0.65    &  26.1       &   787     &    --      &  110  &  \\ % updated 
113 & HIP 108195              &  3.0    & 0.2     &  4.89       &   227     &    --      &   40  &  \\ % updated 
116 & PMM 366328~AB           &  1.82   & 0.56    &  24.0       &  1440     &    --      &  222  &  \\ % updated 
\hline
        \noalign{\smallskip}
         \hline
      \end{tabular}
      \caption*{\footnotesize $^1$Reference number from Tab.~2; $^2$Mass of the inner pair ($M_A + M_B$ from Tab.~2); $^3$Mass of the additional companion; $^4$Outer limit for the stability, calculated using Eq.~\ref{eqn:acrit_cs}}

\end{table*}

\section{Statistical analysis}
\label{sec:main_stat}
\subsection{Statistical formalism}
\label{sec:formalism}

For our statistical analysis we used a Bayesian approach described in \cite{2007ApJ...670.1367L} and in a similar way to what has been done by  \cite{2012AA...544A...9V} and  \cite{2014ApJ...794..159B}.

Our goal is to link the fraction $f$ of the $N$ systems in our sample hosting at least one companion of mass and semi-major axis in the interval $\left[m_{\mathrm{min}},m_{\mathrm{max}}\right] \cap \left[a_{\mathrm{min}},a_{\mathrm{max}}\right]$ with the probability $p$ that such companion would be detected from our observations. 

The likelihood of the data given $f$ is
\begin{equation}
  \label{eq:likelihood}
  L(\{d_j\}|f) = \prod_{j=1}^{N} (1-fp_j)^{1-d_j} \cdot (fp_j)^{d_j}
\end{equation}

\noindent where $(fp_j)$ is the probability of detecting a companion around the jth star, $(1-fp_j)$ is the probability of non detection and $\{d_j\}$ denotes the detections made by the observations, such that $d_j$ equals 1 if at least one companion is detected around star $j$ and 0 otherwise. 

As we have no a priori knowledge of the wide-orbit massive planet frequency, we adopt a {\it maximum ignorance} prior, $p(f) = 1$.
From this prior and the likelihood defined as in Eq.~\ref{eq:likelihood} we can use 
Bayes' theorem to obtain the probability that the fraction of stars having at least one companion is $f$, given our observations $\{d_j\}$, or posterior distribution: 

\begin{equation}
  \label{eq:probability_density}
  p(f|\{d_j\}) = \frac{L(\{d_j\}|f) \cdot p(f)}{\int_0^1 L(\{d_j\}|f) \cdot p(f)df},
\end{equation}

\noindent For a given confidence level $CL =\alpha$ we can then use this posterior distribution $p(f|\{d_j\})$ to determine a confidence interval (CI) for $f$ as follows:

\begin{equation}
  \label{eq:general_ci}
  \alpha = \int_{f_{\mathrm{min}}}^{f_{\mathrm{max}}} p(f|\{d_j\}) df,
\end{equation}

\noindent the boundaries of this CI being the minimal ($f_{\mathrm{min}}$) and maximal ($f_{\mathrm{max}}$) values of $f$ compatible with our observations. 

In case of a null result, clearly $f_{\mathrm{min}} = 0$ and the only result of the such analysis would be a constraint on $f_{\mathrm{max}}$. 

For a case, like ours, where there are some detections, an equal-tail CI can be assumed, and for a given value of $\alpha$, $f_{\mathrm{min}}$ and $f_{\mathrm{max}}$ can be obtained by numerically solving the following equations \citep[see][]{2007ApJ...670.1367L}: 

\begin{eqnarray}
  \label{eq:equal_tail_ci_fmax}
  \frac{1-\alpha}{2} & = & \int_{f_{\mathrm{max}}}^{1} p(f|\{d_j\}) df \\
  \label{eq:equal_tail_ci_fmin}
  \frac{1-\alpha}{2} & = & \int_{0}^{f_{\mathrm{min}}} p(f|\{d_j\}) df
\end{eqnarray}

\subsection{Detection limits}
\label{sec:limits}

For each of the targets in the CBIN sample, we collected the available information on the sensitivity in terms of star/planet contrast at a given angular distance from the star. Such detection limits were therefore used to define the discovery space of our search.
Even if with many common points, the methods used for the evaluations of
the limits are slightly different in the various surveys listed in Tab.~\ref{tab:survey}, the main discriminant being the way in which the noise estimation is made. 

Except for \cite{2005AJ....130.1845L}, which uses a completely different
approach, a Gaussian distribution is assumed for the noise, and a $5-6 \sigma$ level is set for the detection. This is particularly appropriate in case of the ADI data, since the LOCI processing leads to residuals whose distribution closely resembles a Gaussian \citep[see e.g.][]{2007ApJ...670.1367L}. 

\cite{2013ApJ...777..160B} report 95\% completeness levels rather than 5~$\sigma$ thresholds. We therefore used the method described by \cite{2014ApJ...794..159B} to convert them into a common framework with the values from the other studies. % NICI

In the case of the SONG HST survey, 2D detection maps were used.

The COND models \citep{2001ApJ...556..357A,2003AA...402..701B} were used to convert the sensitivity curves into minimum mass limits for all the stars in the CBIN sample.

\subsection{Detection probability}
\label{sec:stat}

In order to evaluate the detection probability $(fp_j)$ for the
targets in our sample, we used the QMESS code \citep{2013PASP..125..849B}. 
The code uses the information on the target stars, together with the detection limits described in Sec.~\ref{sec:limits} to evaluate the probability of detection of companions with semi-major axis up to 1000~au and masses up to 75~$M_\mathrm{Jup}$. 
These values were chosen after a series of tests, aimed at constraining the best possible parameter space for our analysis, given the way our sample was constructed. 

A dedicated version of the QMESS code was used for the target from the SONG HST survey, as 2D contrast maps were provided instead of 1D contrast curves for this purpose \citep[see][for details]{2012AA...537A..67B}. 

In case several limits were available for the same star, separate runs were performed using each limit singularly. 
Then the final detection probability map was built by considering, for each grid point, the highest value among the full set. 
This is equivalent to assume that a planet is detected if it is so in at least one of the images. 

 The same kind of analysis was repeated for the targets in the control SS sample described in Sec.~\ref{sec:ss_sample}. 

Fig.~2 shows the average detection probability map obtained considering all the stars in both the CBIN sample (left panel) and SS control sample (right panel).

\begin{figure*}
\centering
\label{fig:detprob}
\includegraphics[width=8.5cm]{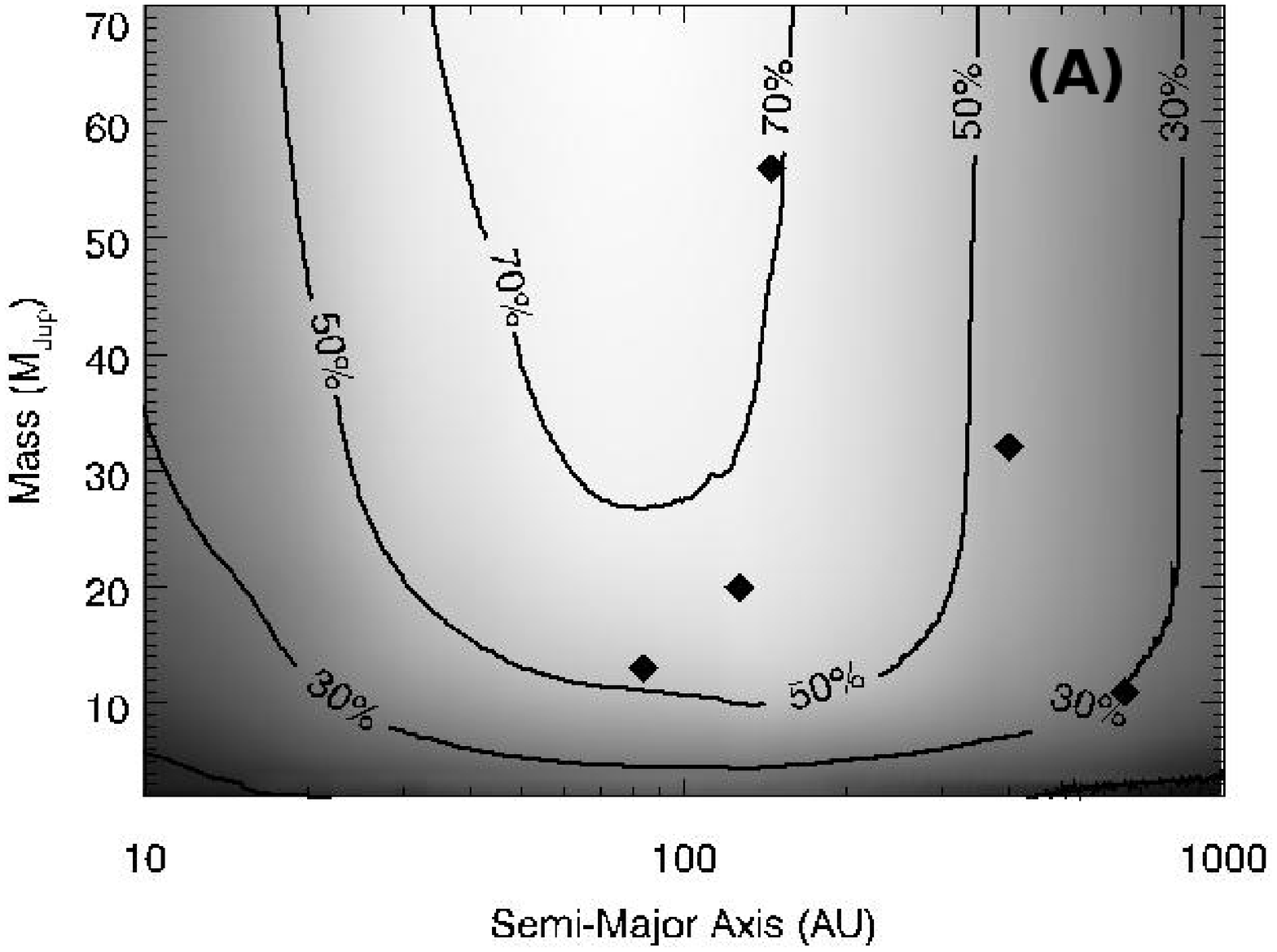}\label{fig:lim_cbin}\hspace{1em}%
\includegraphics[width=8.5cm]{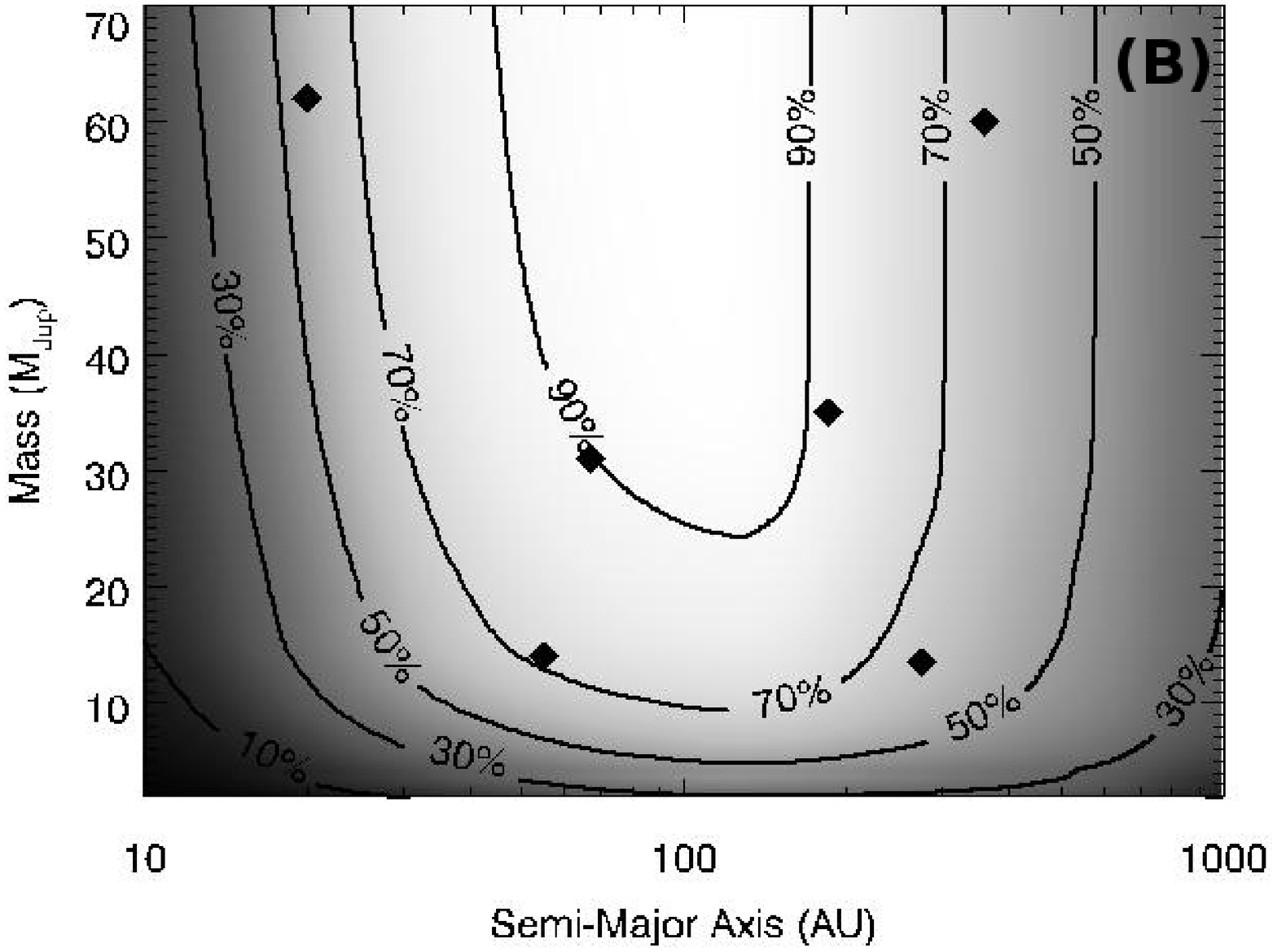}\label{fig:lim_ss}

\caption{Average detection probability as a function of planetary mass and semi-major axis. {\bf A:} Circumbinary (CBIN) Sample; {\bf B:} Comparison (SS) Sample. In both panels the sub-stellar companions reported in Tab.~\ref{tab:dets} are marked with filled diamonds.}
\end{figure*}

\subsection{Derived companion frequency}
\label{sec:cmp_freq}

Five of the 117  systems in the CBIN sample have reported detection of additional sub-stellar companions, two of which (HIP 59960b and 2MASS J01033563-5515561~AB~b) below the deuterium burning limit.
The SS control sample described in Sec.~\ref{sec:ss_sample} includes 7 targets with confirmed sub-stellar companions, including the planetary-mass companions of $\kappa$~And and AB~Pic. The sub-stellar companions HN Peg B \citep{2007ApJ...654..570L} and MN UMa B \citep{2001AJ....121.3235K} are not included in the statistical analysis being at larger projected separation than the limits of the field of view of the imaging surveys considered here.

Table~\ref{tab:dets} summarises the characteristics of the detected companions in both the CBIN and the SS samples.  

We used the approach described in Sec.~\ref{sec:formalism} and the detection probability $(fp_j)$ evaluated as in Sec.~\ref{sec:stat} to constraint the frequency $f$ of sub-stellar companion in wide circumbinary orbits around the targets.  

For a given value $f$ of the fraction of stars having at least one companion in the chosen range of mass and semi-major axis, we inverted Eq.~\ref{eq:general_ci} to estimate its probability $p(f|\{d_j\})$.

Table~\ref{tab:stats} summarises the results we obtained for different choices of mass and semi-major axis ranges, for both the CBIN and the SS sample. Fig.~3 shows the results obtained considering semi-major axis up to 1000~au.

For each case, Eq.~\ref{eq:equal_tail_ci_fmin} and \ref{eq:equal_tail_ci_fmax} were also used to calculate the values of $f_{\mathrm{min}}$ and $f_{\mathrm{max}}$ respectively, for a CL value of 68\% and 95\%.

\begin{table*}
\setcounter{table}{3}
   \caption[]{Sub-stellar companions detections}
     \label{tab:dets}
       \centering
       \begin{tabular}{llcccl}

                      \multicolumn{6}{l}{CBIN Sample} \\
        \hline \hline
        \noalign{\smallskip}
         \#$^1$ & ID            & Mass ($M_\mathrm{Jup}$)   & Sep (au)  & Survey$^2$    & Reference$^3$ \\
        \noalign{\smallskip}
        \hline \hline
             61 & HIP 59960~b   & $11\pm2$                  & $654\pm3$ & JL13          & \cite{2014ApJ...780L...4B} \\
              5 & 2MASS J01033563-5515561~AB~b  & $13\pm1$  & 84        & L15           & \cite{2013AA...553L...5D}  \\
             58 & TWA~5~B       & 20                        & 127       & L05           & \cite{1999ApJ...512L..69L} \\
             22 & HIP 19176 B   & 32                        & 400       & D15           & \cite{2014ApJ...791L..40B} \\
             20 &  H~II~1348~B  & $56\pm3$                  & $145\pm2.3$ & Y13         & \cite{2012ApJ...746...44G} \\
        \hline \hline
        \noalign{\smallskip}
        \multicolumn{6}{l}{SS Comparison Sample} \\
        \noalign{\smallskip}
        \hline \hline
        \noalign{\smallskip}
                & ID            &  Mass ($M_\mathrm{Jup}$)  & Sep (au)  & Survey$^2$    & Reference$^3$ \\
        \noalign{\smallskip}
        \hline
        \hline
               & AB Pic B      & 13.5                & 275             & BN13          & \cite{2005AA...438L..29C}  \\
                & $\kappa$~And~b & $14^{+25}_{-2}$ & $55\pm2$           & B14           & \cite{2013ApJ...763L..32C} \\
                & $\eta$ Tel B  & 20-50               & 185             & BN13          & \cite{2000ApJ...541..390L} \\
                & CD-35 2722~b  & $31\pm8$            & $67\pm4$        & BN13          & \cite{2011ApJ...729..139W} \\
                & HD~23514~b    & $60\pm10$           & 360             & Y13           & \cite{2012ApJ...748...30R} \\
                & PZ~Tel~b      & $62\pm9$            & 20              & BN13          & \cite{2010ApJ...720L..82B} \\
         \noalign{\smallskip}
         \hline \hline
      \end{tabular} 
      \caption*{\footnotesize $^1$Reference number from Tab.~2;  $^2$Original Survey, from Tab.~1;  $^3$Reference for the companion parameters. }
\end{table*}

\begin{table*}[htbp]
\setcounter{table}{4}
   \caption[]{Statistical analysis results.}
     \label{tab:stats}
       \centering
    \noindent\begin{tabular}{cr|cccc|cccc}
        
\multicolumn{2}{r}{} & \multicolumn{4}{c}{CBIN Sample}   & \multicolumn{4}{c}{SS Comparison Sample}\\ 

\hline \hline
\noalign{\smallskip}
\noalign{\smallskip}
SMA  &  Mass & $N_{det}^1$ & $f_{best}^2$  & \multicolumn{2}{c|}{$[f_{min}, f_{max}]^3$} & $N_{det}^1$ & $f_{best}^2$  & \multicolumn{2}{c}{$[f_{min}, f_{max}]^3$}   \\
(au) &  ($M_\mathrm{Jup}$) &               &    (\%)       &  CL=68\%   & CL=95\%       &            &  (\%)         &  CL=68\%             & CL=95\%            \\
 \noalign{\smallskip}
 \hline \hline
         \noalign{\smallskip}
\noalign{\smallskip}

 $10-100$  &  $ 2-15$      &  1 & 1.35  & [0.95,  4.30]  & [0.35,  7.20]  &  1 & 0.90  & [0.65,  2.85]  & [0.25,  4.80] \\
           &  $15-70$      &  0 & --    & [0.00,  1.95]  & [0.00,  3.85] &  2 & 1.20  & [0.85,  2.70]  & [0.40,  4.20] \\
           &  $ 2-70$      &  1 & 1.15  & [0.80,  3.60]  & [0.30,  6.05] &  3 & 1.90  & [1.35,  3.70]  & [0.70,  5.45] \\
           \noalign{\smallskip}

           \hline
           \noalign{\smallskip}
  $10-500$ &  $ 2-15$      &  1 & 1.30  & [0.95,  4.10]  & [0.35,  6.85]  &  2 & 1.60  & [1.10,  3.60]  & [0.50,  5.60] \\
           &  $15-70$      &  3 & 3.30  & [2.30,  6.30]  & [1.20,  9.25] &  4 & 2.50  & [1.80,  4.40]  & [1.05,  6.25] \\ 
           &  $ 2-70$      &  4 & 4.50  & [3.20, 7.80]  & [1.85, 11.00] &  6 & 3.95  & [2.95,  6.15]  & [1.90,  8.35] \\
           \noalign{\smallskip}
           \hline
           \noalign{\smallskip}
 $10-1000$ &  $ 2-15$      &  2 & 2.70  & [1.85,  6.00]  & [0.85,  9.25] &  2 & 1.85  & [1.30,  4.20]  & [0.60,  6.55] \\
           &  $15-70$      &  3 & 3.55  & [2.50,  6.75]  & [1.30,  9.90] &  4 & 3.05  & [2.20,  5.30]  & [1.25,  7.55] \\
           &  $ 2-70$      &  5 & 6.00  & [4.35,  9.75]  & [2.70, 13.35] &  6 & 4.70  & [3.50, 7.35]  & [2.25,  9.95] \\   

           \noalign{\smallskip}
         \hline \hline
         \noalign{\smallskip}
 \noalign{\smallskip}
      \end{tabular} 
      \caption*{\footnotesize $^1$Number of detections in the considered mass and semi-major axis (SMA) range; $^2$Best value of the planet frequency compatible with the observations; $^3$Minimum and maximum values of the frequency compatible with the results, for a given confidence level (CL). }
\end{table*}

\begin{figure*}[htbp]
\label{fig:freq}
\centering
\label{fig:plfreq_cbin}\includegraphics[width=8cm]{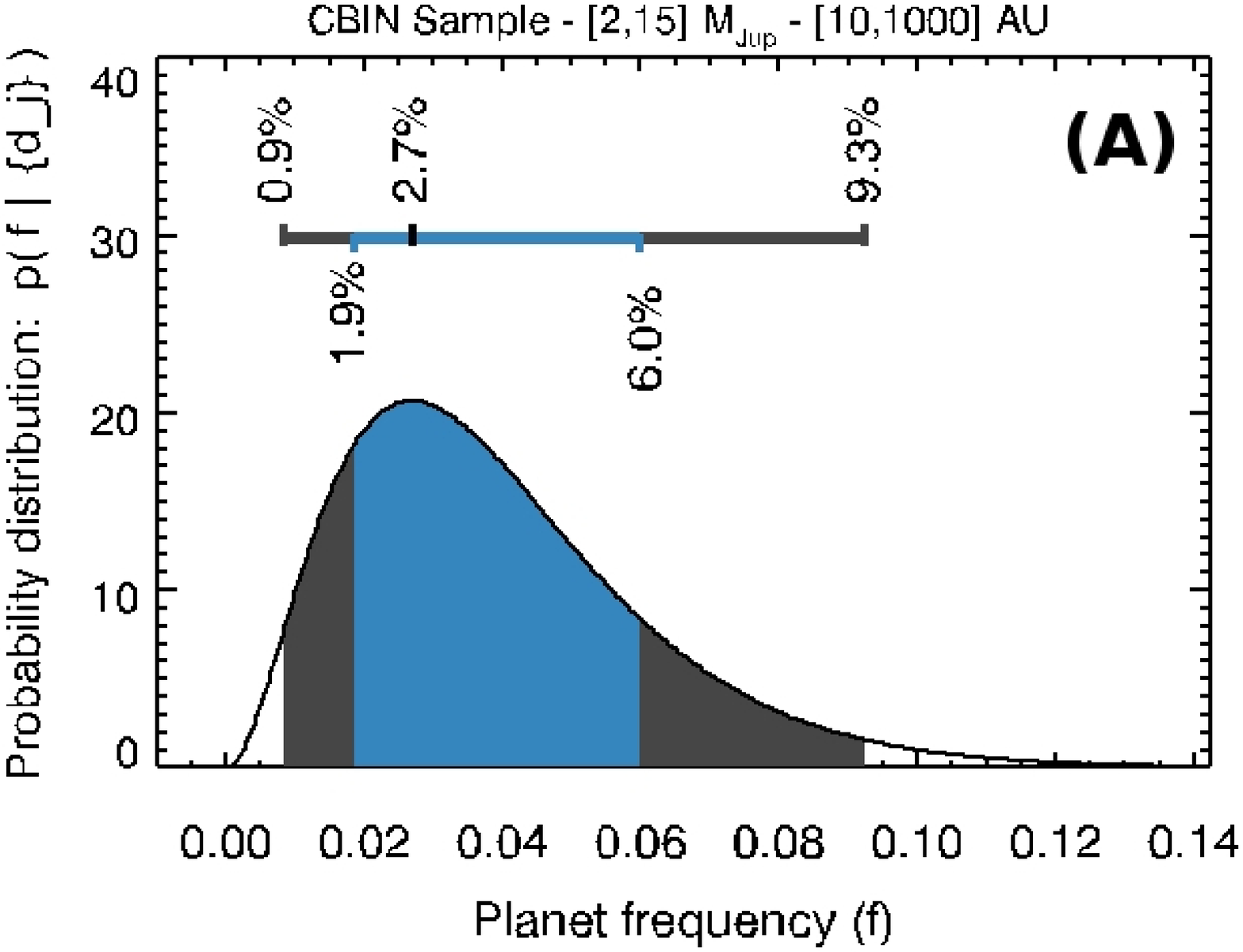}\hspace{0.1em}%
\label{fig:freq_cbin}\includegraphics[width=8cm]{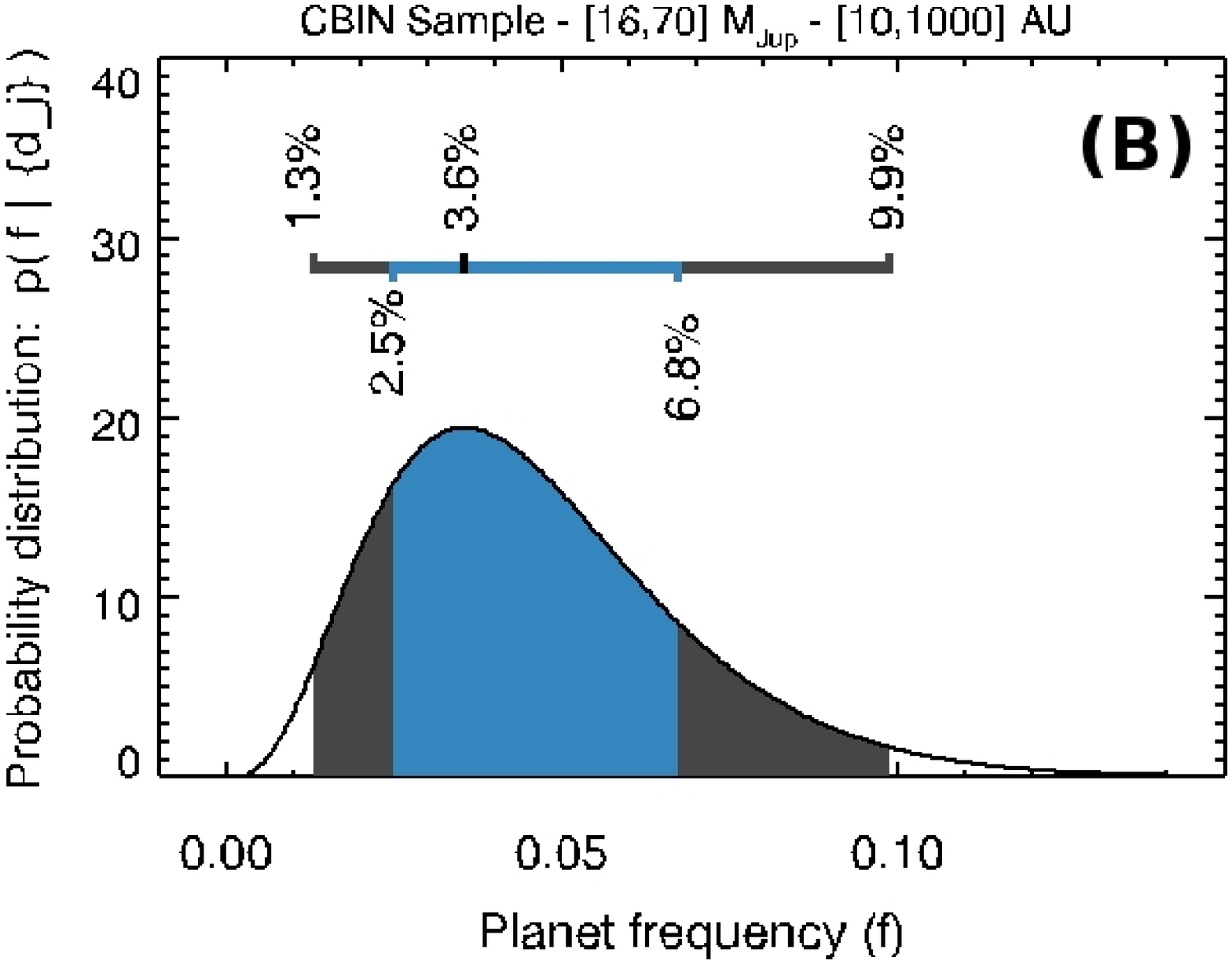}
\label{fig:plfreq_ss}\includegraphics[width=8cm]{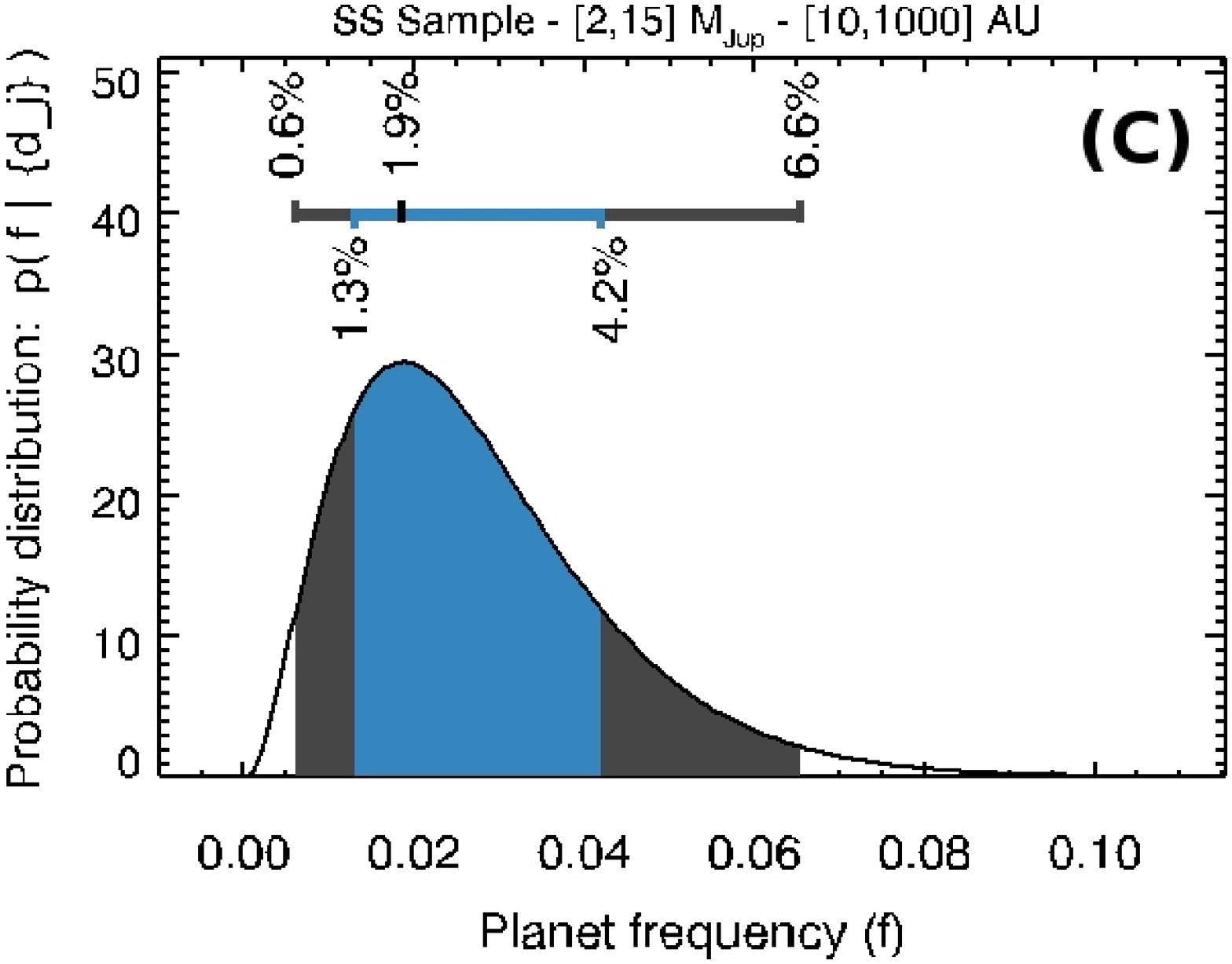}\hspace{0.1em}%
\label{fig:freq_ss}\includegraphics[width=8cm]{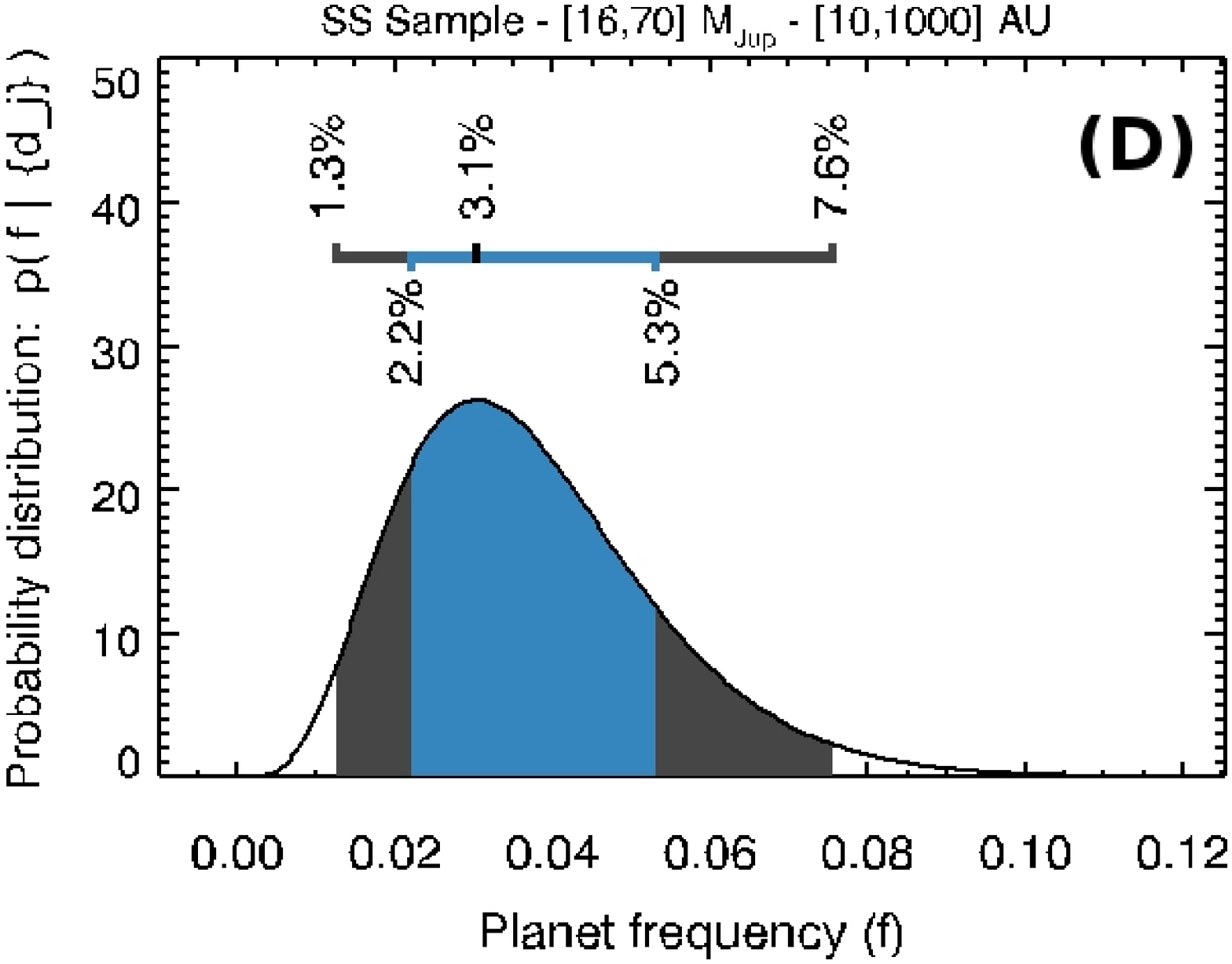}
\caption{Probability distribution (see Tab.~\ref{tab:stats} for details) of the frequency of planetary mass (up to 15~$M_{Jup}$, left panels) and BD (16-70~$M_{Jup}$, right panels). The results for the CBIN sample and the and the SS control sample are shown in the upper and lower panels, respectively. 
The shaded areas show the frequency limits for the 95\% (gray) and 68\% (blue) confidence levels.}
\end{figure*}

\section{Discussion}
\label{sec:discussion}

\subsection{The frequency of planets and brown dwarfs in circumbinary orbits}

We have presented the results of the statistical analysis of a sample of 117 tight binaries observed in the contest of some of the deepest DI planet search surveys. 
Five of the targets included in our sample have reported detection of sub-stellar companions, two of which (HIP 59960~b and 2MASS~J01033563-5515561~b) are in the planetary mass regime.

We find that our data are compatible with 6\% (with an upper limit of $\sim$13\% at 95\% confidence level) of tight binaries hosting sub-stellar companions ($ 2~M_{Jup} < M_c < 70~M_{Jup}$) within 1000~au. 
If we limit our analysis to planetary mass companions ($ 2~M_{Jup} < M_c < 15~M_{Jup}$), the best frequency value is 2.70\%  (with a 95\% CL upper limit of $\sim$9\%), for a semi-major axis cut-off of 1000~au, and 1.35\% (with $f_{\mathrm{max}}\sim~$7\% at 95\% CL) for separations up to 100~au. 

From a similar analysis of the SS control sample described in Sec.~\ref{sec:ss_sample} we were able to infer a frequency of companions within 1000~au between 0.6\% and 6.55\% for the planetary mass objects and between 2.25\% and 9.95\% for companions up to 70~$M_{Jup}$,  within the same semi-major axis range. 

Although our results seem to point towards the existence of small differences between the frequency of sub-stellar companions around close binaries and isolated stars, the significance of such result is only marginal (at most 2$\sigma$ for the 1000~au case, as shown also in Fig.~3) and needs confirmation through deeper observations and using larger samples. 

Furthermore, possible selection effects may play a role.
In particular, the discovery of substellar objects around a target may have triggered dedicated follow-up observations resulting in improved sensitivity to very close stellar companions.
This is likely the case of HIP 59960, while the other stellar companions of stars in Table~4 were known in advance or presented in the discovery papers of the substellar companions.
Our results therefore seem to suggest that no strong difference exists, in terms of frequency of sub-stellar companions in wide orbit, between close binaries and single stars. 

\subsection{Comparison with Kepler results}

\citet{2012Natur.481..475W} estimated a frequency or circumbinary planets of about 3\% (with lower limit of 1\%) when
considering the short-period circumbinary planets detectable by Kepler.
The separation range we are sensitive with direct imaging is different from that explored by Kepler and then the two techniques are highly complementary. Very recently, a circumbinary planet at 2.7~au was discovered with Kepler
\citep{2015arXiv151200189K}, indicating that circumbinary planets likely are present over an extended separation range.

Some additional interesting trends are also emerging from the Kepler sample.
\cite{2014IAUS..293..125W} noticed the complete absence of transiting circumbinary planets around binaries with p~$<$~5~d. This seems unlikely to be due to selection effects. Indeed, according to \cite{2011AJ....142..160S} a relatively high number of these systems were in fact observed by Kepler. Moreover, such planets, as long as they are near the inner stability limit, would have an higher transit probability, and therefore be easy to detect. 
The lack of planets around very close binaries could be due to the formation history of the tight pair, which may be linked to the presence of an outer stellar companions which shrunk the central binary orbit via Kozai mechanism and tidal circularization \citep{2015MNRAS.453.3554M}.
\cite{2014AA...570A..50S} suggest strong photoevaporation, expected for this kind of tight binaries which keep fast rotation and high levels of magnetic activity for their whole lifetime, as a possible explanation for this lack of planets. 

Our sample includes a large variety of binary configuration, with a fraction of binaries with very short periods (17\%),a number of binaries with orbital periods comparable to those of the hosts of Kepler circumbinary planets (7-41 days) and a significant number of wider binaries.
Therefore, the possible lack of planets around very close pairs due to dynamical interaction has not a dominant role in our statistical analysis.
Unfortunately, the binary properties of systems with detected sub-stellar companions are poorly constrained (orbits not available)
for HIP 19176, HII 1348, 2MASS~J01033563-5515561, and HIP 59960, while a reliable orbital solution was derived for TWA 5. However, a very close system is possible only for HII 1348.

Another property emerging from Kepler results is that often the circumbinary planets are found close to the dynamical stability limits. This is likely due to stopping of inward migration close to the inner disk limits caused by the presence of central binary \citep[see e.g.][]{2013AA...556A.134P}.
The circumbinary sub-stellar objects identified with direct imaging are typically very far from the dynamical stability limits with only 2MASS~J01033563-5515561~b  being at a separation which is less than two times the adopted dynamical stability limit. This holds both for the objects included in the sample as well as for other circumbinary planets or brown dwarfs which are not included in our statistical analysis due to the lack of suitable publication of the parent sample such as Ross 458 \citep{2010ApJ...725.1405B} and FW Tau and ROX42B \citep{2014ApJ...781...20K}  or because the binary is wider than our adopted limit, as 
SR12 \citep{2011AJ....141..119K}.
This could be explained by a different formation mechanism but ejection to outer orbits due to gravitational encounters is also a viable possibility. The system around HIP 59960 is of special interest in this context, thanks to the presence of both a circumbinary companion of planetary mass at wide separation and of a circumbinary disk which have been recently spatially resolved with SPHERE and GPI \citep{2016A&A...586L...8L,2015ApJ...814...32K}.
The on-going extension of the SPOTS program with SPHERE at VLT, probing closer separations, will be crucial
for a better understanding of the separation distribution of circumbinary sub-stellar objects.

\subsection{Implications for the origin of planet candidates around post-common envelope binaries}

In the past years, several claims of massive planetary companions orbiting post-common envelope binaries,
based on the transit timing technique, appeared in the literature \citep[][and references therein]{2013AA...549A..95Z}. 
Their existence is currently controversial, as in several cases the continuation of the observations did not follow the ephemeris
from the discovery papers, calling for a full revision of the orbital elements and/or the inclusion of additional objects 
\citep[see,  e.g. ][]{2010MNRAS.407.2362P, 2012AA...543A.138B}. In other cases, the
proposed multi-planet systems are not dynamically stable \citep[see, e.g., ][]{2013MNRAS.435.2033H}.
Only the system orbiting NN Ser appears to be confirmed \citep{2014MNRAS.438L..91P}, as timing variations
are consistent with circumbinary planets for both the primary and secondary eclipses. 
The recent imaging non-detection of the brown dwarf candidate identified with timing technique around
V471 Tau \citep{2015ApJ...800L..24H} further calls into question the Keplerian origin of the observed eclipse timing variation
\citep[see however ][ for a different interpretation of the imaging non-detection]{2015ApJ...810..157V}.

If the observed timing variations are due to circumbinary planets, there are two paths for their formation.
The first one is that they formed together with the central binary and survived the common envelope evolution
of central pair (first generation scenario). In most cases, the observed wide separation could be compatible with this possibility.
The second scenario is that circumbinary planets formed after the common envelope evolution, in the circumbinary disk that is expected to form from the material lost in the process. The large content of heavy elements expected in such disks \citep{1998Natur.391..868W} could contribute in a large efficiency of planet formation process in these environments. This scenario is favoured in the discussion by \citet{2013AA...549A..95Z} and, for the specific case of the NN Ser system, by \citet{2013MNRAS.436.2515M},
while \citet{2014MNRAS.444.1698B} identified some difficulties with the second-generation model. 

The first attempt to estimate the frequency of circumbinary planets around post-common envelope
binaries was performed by \citet{2013AA...549A..95Z}. They found a very high frequency (90\% from 10 systems with adequate time baseline and measurement accuracy) of the occurrence of eclipse timing variations suggesting the presence of circumbinary planets. 
In most cases, these candidate companions are moderately massive (5-10 $M_{J}$)
and at moderately wide separation (5-10~au), i.e. within the mass and separation range we
are probing with direct imaging (although the binary evolution could have caused some outward migration due to system mass loss).
The similar (and relatively low) frequency of sub-stellar objects around close binaries and single stars found in our work points against the first-generation scenario being responsible for the majority of planet candidates around post-common-envelope binaries. 
This leaves as the most probable interpretations to the eclipsing timing variations either second generation
planet formation or some non-Keplerian physical mechanisms mimicking the timing signature of planetary
companions.
It should be noticed that second generation planets are expected to be much younger than the age of the
system and thus significantly brighter than 1st generation ones. 
This would strongly favour their direct detection. In the case of the NN Ser system, the cooling age of the white dwarf in the system is estimated to be just 1 Myr \citep{2010AA...521L..60B}.
We note that in the three cases of post-common envelope systems in our sample (Algol, Regulus, $\theta$ Hya),
the detection limits were derived for the original system age, and thus are valid for first generation planets.
 Lower mass limits could be derived for planets formed at the time of the common-envelope evolution.

\section{Summary and conclusions}
We have presented a statistical analysis of the combined body of 
existing high-contrast imaging constraints on circumbinary planets, to complement our ongoing  SPOTS direct imaging survey dedicated to such planets.
The sample of stars considered includes 117 objects and comes from a search for tight binaries within the target lists of 23 published direct imaging surveys, including some of the deepest ones performed to data. This resulted in a large variety of binary configurations, including systems with very short periods, a number of binaries with orbital periods comparable to those of the hosts of Kepler circumbinary planets and a significant number of wider binaries.

The main conclusion of this work is the suggestion that no strong difference exists, in terms of frequency of sub-stellar companions in wide orbit, between close binaries and single stars.

With five of the pairs included in our circumbinary sample hosting sub-stellar companions, only two of which have planetary mass, we were able to constraint the frequency of circumbinary companions in wide orbits ($<$~1000~au) to a value between 
$\sim$0.9\% and $\sim$9\% for the planetary mass companions, and between 1.3\% and $\sim$10\% for low-mass brown dwarfs, with a confidence level of 95\%.

A similar analysis for the comparison sample of 205 single stars lead to a value of the frequency of planetary (low-mass BD) companions between 0.6\% and 6.55\% (1.25\% and 7.55\%), with the same confidence level. 

Although there seem to be some small differences between the results for the two samples, the retrieved values of the frequency are compatible within the errors, and given the small number of target considered, it is premature to speculate about possible differences in the overall frequency, as well as in the formation mechanisms.

The similar (and relatively low) frequency of sub-stellar objects around close binaries and single stars also points against the first-generation scenario being responsible for the high abundance of planet candidates around post-common-envelope binaries.

This leaves as the most probable interpretations to the eclipsing timing variations observed in the majority of post-common envelope binaries either second generation planet formation or some non-Keplerian physical mechanisms mimicking the timing signature of planetary companions.

Our result nicely complement those coming from the Kepler spacecraft, as the separation range explored with direct imaging is quite different. Kepler's circumbinary planets are often close to the dynamical stability limit, whereas most the companions identified with direct imaging are instead much further out. 

The on-going extension of the SPOTS program with SPHERE at VLT, probing closer separations, will be crucial for a better understanding of the separation distribution of circumbinary sub-stellar objects.

\bibliographystyle{aa}
\bibliography{MBonavita_SPOTS_II_final}

\subsubsection*{Acknowledgements}

This research has made use of the SIMBAD database
and of the VizieR catalogue access tool operated at CDS, Strasbourg, France, and of the Washington Double Star Catalogue maintained at the U.S. Naval Observatory.
We made use of data products retrieved from ESO Science Archive Facility (program 084.A-9004) and SOPHIE archive.
The authors warmly thank Anne-Marie Lagrange for sharing results in advance of publication.
The authors would like to thank Beth Biller, Brendan Bowler, Alexis Brandeker, Sebastian Daemgen, Philippe Delorme, Markus Kasper, David Lafreniere, Anne-Lise Maire and Inseok Song for providing the detection limits and the target information from their surveys.
We thank the anonymous referee for extensive feedback that significantly improved the clarity of the paper.
SD acknowledges support from  the  “Progetti  Premiali”  funding  scheme  of  the  Italian  Ministry of Education, University, and Research.

\clearpage

\onecolumn
\include{mastertable}

\twocolumn

\appendix

\section{\bf Notes on individual objects}
\label{app:notes}

%\subsection{$acrit < 20$} 

\begin{enumerate}
    \item {\bf TYC 5839-0596-1} %% 1
    See \cite{2015AA...573A.126D}

\item {\bf  HIP 3210}  %% 2
Classified as SB2 from \cite{2004AA...418..989N} with mass ratio of 0.35.
The orbital solution is not available.
\cite{2013MNRAS.435.1376M} classified the star as a member of Columba association
on the basis of the strength of lithium and kinematics, but without taking multiplicity
into account. The system is also moderately X-ray bright.
Considering the limited sensitivity of these age indicators for late F stars and the complications
introduced by multiplicity, we adopt an age of 150 Myr.
The possibility of tidal locking can not be ruled out but the
young disk kinematics and lithium would make an old age unlikely.
The confirmation of Columba membership would require additional data on binary orbital solution.

\item {\bf HIP 3924}  %%3
See \cite{2015AA...573A.126D}

\item {\bf HIP 4448 = HD 5578 = BW Phe} %%4
Classified as a new potential member of Tuc-Hor association in
\cite{2001ApJ...559..388Z} \citep[but not included in the list by][]{2004ARAA..42..685Z}.
\cite{2008arXiv0808.3362T} and \cite{2013ApJ...762...88M} 
instead classified it a member of Argus association.
The age indicators support a young age with upper limit of 150 Myr.
We then adopt Argus membership and age, but stressing the uncertainty
in the kinematic parameters due the unknown binary orbit.
Indeed, the star is a close binary with similar components (projected separation
0.228 arcsec)  

\item {\bf 2MASS J01033563-5515561} %%5
Close visual pair with a detected companion close to deuterium burning mass
in circumbinary configuration \citep{2013AA...553L...5D}.
The system is a probable member of Tuc-Hor association.

\item {\bf HIP 4967 = G 132-50A} %%6
Young M dwarf, probable member of AB Dor MG, resolved into a tight binary by B15.
There is an additional wide companion G 132-50B at 25.6 arcsec, which
it itself a 2 arcsec pair, making the system quadruple.
%% masses of B 0.63+0.59 Janson astralux

\item {\bf HIP 9141 = HD 12039 = DK Cet}  %%7
Member of Tucana association.
A close stellar companion ($\rho$ = 0.15 arcsec) was 
imaged by \cite{2007ApJS..173..143B}. 

\item {\bf NLTT 6549} %%8
Young M dwarf, possible member of Hyades stream, resolved into a tight binary by B15.
We adopt the parameters by B15.

\item {\bf HIP 11072 = HD 14802 = $\kappa$ For}  %%9
Triple system, formed by a solar type star and a close pair of M dwarfs with
tentative period of about 3 days. A full orbital solution of the outer orbit is
available \citep{2013AJ....145...76T}, including RV, astrometry and resolved 
imaging of the components.
Isochrone fitting from \cite{2009AA...501..941H} gives $5.7\pm0.5$ Gyr, fully 
consistent with the low chromospheric emission reported from 
\cite{2004ApJS..152..261W} ($\log R_{HK}=-5.05$). The X ray emission is instead larger, 
comparable to Hyades stars of similar colour, but this may be dominated by
the emission from the close pair of M dwarfs due to their probable tidal locking.
\cite{2007ApJ...669.1167B} report a gyro-age of 730 Myr, from a rotation 
period of 9 days, that is wrong due to a typo in \cite{2003AA...397..147P}
\citep[the referenced paper][gives 19.3 days, derived from chromospheric 
emission]{1997AA...326..741S}. We then adopt the isochrone age.

\item {\bf HIP 12413 = HD 16754A = s Eri } %%10
Star with various signatures of multiplicity.
As discussed in \cite{2011ApJ...732...61Z}, the high-resolution X-ray imaging by
\cite{2007AA...475..677S} indicates that the early-type primary should have a 
spatially unresolved low mass companion.
The presence of RV variations \citep{1961MNRAS.123..233B} and of the 
astrometric acceleration in Hipparcos catalogue further
support the binarity and suggest an orbital period of several years.
We derive the stability limit for a semi-major axis of 5~au
and a mass of 0.6 $M_{\odot}$.
There is an additional M-type companion at 24 arcsec.
The system is a probable member of Columba association \citep{2011ApJ...732...61Z}.

\item {\bf HIP 12545 = BD +05 0378}  %%11
See \cite{thalmann14}
Member of BPIC MG. Identified as SB1 in \cite{2003ApJ...599..342S} (peak-to-valley variation of 20 km/s,
no orbital solution provided). However, \cite{2012ApJ...749...16B} found no evidence for large RV variations from their monitoring over 600 days (14 epochs, scatter of 179 m/s).

\item {\bf HIP 12638 = HD 16760} %%12
Radial velocity monitoring revealed a sub-stellar companion of projected mass
$m \sin{i}$ about $14~M_{J}$ \citep{2009ApJ...703..671S,2009AA...505..853B}. The direct detection by \citet{2012ApJ...744..120E} shows that
the true mass is significantly larger than the minimum mass and that the inclination is very close
to pole-on.
\citet{2012ApJ...744..120E} derived a combined imaging and RV orbital solution, which we adopt in our study.
\citet{2012ApJ...744..120E} also summarised the puzzling results from different age diagnostics.
The adopted age is derived from the membership to AB Dor moving group.
The star has a wide companion (HIP 12635) at 14 arcsec.

\item {\bf HIP 13081 = HD 17382 = BC Ari = GJ 113}   %%13
Triple system. 
The primary is a spectroscopic and astrometric binary (Hipparcos acceleration).
\cite{2002AJ....124.1144L} derived a preliminary spectroscopic orbital solution 
with period about 17 yr in a rather eccentric orbit. 
The minimum mass of the companion is about $0.18~M_{\odot}$.
There is also a wide companion (GJ 113 C) at 20 arcsec (mass $M_B=0.16 M_{\odot}$). 
The star is a probable member of Hercules-Lyra according to 
\cite{2008MNRAS.384..173F}.
Activity indicators are consistent with a slightly older age (about 400 Myr) while
lithium was not detected in the spectrum \citep{1996AA...311..951F} 
suggesting an age of about 600 Myr or older.
We then consider the membership unlikely, as also concluded by \cite{2013AA...556A..53E}.
The discrepancy between age indicators might also be explained if the unseen companion
is actually white dwarf rather than a low mass main sequence star 
\citep[see ][for the case of HD8049]{2013AA...554A..21Z}.
But considering the lack of evidences supporting this latter hypothesis and the
marginal amount of the discrepancy between age derived from lithium and activity
indicators, we adopt an age 500 Myr.

\item {\bf HIP 14555 = GJ 1054 A} %%14
Short-period SB2 with similar components. 
See  \citet{2014AA...566A.126M}.

\item {\bf HIP 14576 = Algol = HD 19356}  %%15
Triple system, with an inner pair evolved through mass transfer phase, and an additional
component that is anyway close enough (a=2.78~au) to allow the search for planets around
the three stars. Stellar masses and orbital parameters from \cite{1993MNRAS.262..534S}.

\item {\bf HIP 16247 = HD 21703 = AK For}  %%16
Eclipsing binary recently studied by \cite{2014AA...567A..64H}.
The high levels of chromospheric and coronal activity are due to tidal locking and not to young age,
as indicated by the lack of detection of lithium by \cite{1995AA...295..147F}, that 
corresponds to a lower limit to stellar age of about 200 Myr.
The thin disk kinematics is compatible with an age similar to that of the Sun.

\item {\bf 2MASS J03363144-2619578 = SCR J0336-2619} %%17
New close visual binary from Lannier et al. 2015; probable member of
Tuc-Hor or Columba associations according to \citet{2013ApJ...774..101R}.

\item {\bf  HIP 16853 = HD  22705}  %%18
See \cite{thalmann14}

\item {\bf HD 282954} %%19
SB2 in Pleiades open cluster according to 
\cite{1998AA...335..183Q}. No orbit available.

\item {\bf HII 1348}  %%20
SB2 in Pleiades open cluster according to \cite{1998AA...335..183Q}. No orbit available
Individual masses 0.67 and 0.55 $M_{\odot}$ from \cite{2012ApJ...746...44G}
Circumbinary brown dwarf detected by \cite{2012ApJ...746...44G} and
\cite{2013PASJ...65...90Y}.

\item {\bf HD 23863} %%21
Close visual companion in Pleiades open cluster detected by 
\cite{2012AA...541A..96R} using the lunar occultation technique at a projected 
separation of 22.1 mas=2.95~au.
Estimated individual magnitudes are 7.60  and 1.66 in K band, that, coupled with the distance
and age of the Pleiades, lead to individual masses of $1.75$ and $0.45~M_{\odot}$.
The star is also a SB according to \cite{1991ApJ...377..141L}
\cite{2012AA...541A..96R}  were not able to conclude whether this is the same object
responsible of the RV variations, due to the scarcity of the available info
on the RV variations. 

\item {\bf HIP 19176 = HD 284149} %%22
A brown dwarf companion was recently detected by \citet{2014ApJ...791L..40B} 
at a projected separation of about 400~au.
As discussed in this paper, the RV variability indicates the presence
of an additional companion at small separation.
We adopt the stellar parameters from \citet{2014ApJ...791L..40B}.

\item {\bf RX J0415.8+3100 = V952 Per }   %%23
This star was classified as a short-period SB1 
by \cite{2012ApJ...745..119N} on the basis of the
large (70 km/s) RV variations over timescales of days.
A lower limit to the companion mass is $0.21~M_{\odot}$ 
assuming a period of 2 days and a RV semi-amplitude of 35 km/s.
An additional component at 0.9 arcsec makes the system triple.
\cite{2015ApJ...799..155D} classified the star as member of the 
Taurus Extended association.
We estimated a distance of 200 pc with a reddening E(B-V)=0.15, after correcting the system magnitude
for the presence of the visual companion and assuming negligible flux contribution by the 
spectroscopic component.
An age of about 100 Myr is estimated from Lithium EW.

\item {\bf RX J0435.9+2352 = V1324 Tau} %%24
Close visual binary. D15 classified it
in the Taurus extended group.
We adopt an age of 20 Myr following D15.

\item {\bf HIP 21482 = HD 283750 = V833 Tau}  %%25
Triple system: V833 Tau is a spectroscopic binary with
period 1.79 days. The mass of the companion has been estimate by \cite{2008MNRAS.384..173F}
to be 0.19 $M_{\odot}$, leading to a total mass of V833 Tau Aab of 1.03 $M_{\odot}$.
The system has a wide companion (WD0433+270) at 126 arcsec.
The primary appears to be a member of Hyades group. However,
the WD cooling age is not compatible with 
the Hyades age unless the rather exotic scenario
of a Fe-core WD favoured by \cite{2008AA...477..901C}.
Following \cite{2008AA...477..901C} we then adopt the Hyades 
age (625 Myr) but a much older age (about 4 Gyr) can not be
ruled out. The high metallicity 
is compatible with Hyades membership.

\item {\bf GJ 3305} %%26
Member of $\beta$ Pic  MG. Close visual binary discovered by K07. 
An orbital solution was derived by \cite{2012AA...539A..72D}.
The pair has also a wide companion \citep[sep. 66 arcsec, see][]{2006AJ....131.1730F}, 
the F0V star 51 Eri.

\item {\bf HIP 21965 = HD 30051}   %%28
Astrometric binary, with orbital solution derived by \citet{2007ApJS..173..137G}.
The star is a member of Tuc-Hor association.

\item {\bf DQ Tau}  %%29
SB2 with nearly identical components, member of Taurus star forming region.
Orbital parameters from \cite{1997AJ....113.1841M} and primary mass from \cite{2015ApJ...799..155D}.

\item {\bf HIP 23296 = HD 32115}  %%30
This is a slow rotating A type star without abundance anomalies.
It is a short-period single-lined SB with orbital parameters derived in \cite{2006AJ....132.1490F}.
The minimum mass is of $0.29~M_{\odot}$ (for a primary stellar mass of $1.5 ~M_{\odot}$)
A very low mass star in wide orbit has been identified by \cite{2014MNRAS.437.1216D}.
V12 adopt an age of 125 Myr from the position on CMD similar to
Pleiades stars.

\item {\bf HIP 23418 = GJ 3322 = 2MASS J05015881+0958587 } %%31
Tight triple system, formed by a 12d spectroscopic binary and an outer visual companion
at 1.37 arcsec that strongly limits the region allowed for stable circumbinary planets around
the central pair.
We adopt the trigonometric distance from \citet{2014AJ....147...85R}, the age from membership
to $\beta$ Pic MG and masses from \citet{2008MNRAS.389..925T}.

\item {\bf L449-1AB:}   %%32
See \cite{2015ApJS..216....7B}.

\item {\bf HIP 25486 = HD 35850 = AF Lep:}   %%33
See \cite{thalmann14}

\item {\bf AB Dor AC = HIP 25647 AC = HD 37065 AC} %%34
First of the two close pairs in the AB Dor quadruple system.
Resolved by \cite{2005Natur.433..286C}. Astrometric orbit
has been derived by \cite{2006AA...446..733G}. We adopt these parameters in our analysis.
The secondary AB Dor C is a very low mass star ($0.09 M_{\odot}$). Included in the B07 survey.

\item {\bf AB Dor BaBb = HIP 25647 BaBb = HD 37065 BaBb} %%35
Second pair in the AB Dor quadruple system.
Resolved into a 0.06 arcsec binary by \cite{2007AA...462..615J}, included in the CH10 survey. 

\item {\bf 2MASS J05320450-0305291 =  V1311 Ori = TYC 4770-797-1}  %%36
Close visual binary, member of $\beta$ Pic MG.
Individual masses from \citet{2012ApJ...754...44J} and distance from L15.

\item {\bf HIP 30920~A = GJ 234 A = V575 Mon}  %%37
Spectroscopic, astrometric and visual binary.
Parameters from \cite{2000AA...364..665S}
The stellar age is uncertain but likely moderately young, considering
the large X-ray emission, significant rotation and young disk kinematics.
We adopt 150 Myr.

%%% NEW Brandt et al. 2013
\item {\bf HIP 32104 = HD 48097 = 26 Gem  = HR 2466} %%38
Member of Columba association according to \cite{2011ApJ...732...61Z} and  \cite{2013ApJ...762...88M}.
Spectroscopic \citep{2005AA...443..337G} and astrometric (Hipparcos orbital solution) binary.
Combining the spectroscopic solution with the inclination from Hipparcos results
in a companion mass of $0.51 M_{\odot}$ at 1.87~au.
The secondary is most likely responsible for the X-ray emission from the system.

\item {\bf HIP 35564} See \cite{2015AA...573A.126D}. %%39

\item {\bf HIP 36349 = V372 Pup = 2MASS J07285137-3014490 = GJ 2060} %%40
Close visual system member of the AB Dor MG.

\item {\bf HIP 36414} See \cite{2015AA...573A.126D}. %%41

\item {\bf GJ 278C = YY Gem = Castor C}   %%42
Eclipsing binary with similar components \citep[P=0.81d, 
M=0.5975+0.6009, ][]{2002ApJ...567.1140T}.
The other components of the Castor system (two SB with A type primaries) are
at 72 arcsec = 1070~au \citep[total mass 4.83 $M_{\odot}$][]{2002ApJ...567.1140T}.
Distance to the system from \cite{2002ApJ...567.1140T}, based on reanalysis of Hipparccos data.
Member of Castor MG \citep{2003AA...400..297R}. % (age 320 Myr)

\item {\bf HIP 38160 = HD 64185}  %%43
A close visual companion at 0.141 arcsec=4.8~au has been reported by R13. 
This companion might also be responsible of the astrometric signature in \cite{2005AJ....129.2420M}.
We adopt the mass of the companion from R13.
The star listed in CCDM and WDS (CCDM J07492-6017B) at a projected separation
of 23 arcsec is not physically associated.
The star is a member of Carina-Near MG according to \cite{2006ApJ...649L.115Z}. 

\item {\bf HIP 39896~A = FP Cnc = GJ 1108A}   %%44
The star is a probable member of Columba association according to B13.
They also discovered a close visual companion (sep 0.25 arcsec).

\item {\bf HIP 39896~B = GJ 1108B} %%45
Additional close pair (SB2) of M dwarf companions at a separation of 14 arcseconds from HIP 39896~A 
We adopt the discovery parameters by \cite{2012ApJ...758...56S}.
Both pairs have been observed in deep imaging.
There is a limited space of dynamical stability (from 23 to 68 au) for planets 
around the central binary, due to moderately wide orbit of the central binary
and the presence of the outer pair.

\item {\bf EM Cha = RECX7}   %%46
See \cite{thalmann14}.

\item {\bf RS Cha = HIP 42794 = RECX8}  %%47
SB2 and EB with similar components, member of $\eta$ Cha open cluster
See \cite{2005AA...442..993A} and references therein for a detailed
description of the system. 
One of the components is also a pulsating $\delta$ Scu star

\item {\bf EQ Cha = RECX12}   %%48
Close visual binary member of $\eta$ Cha open cluster (B06).
Flux ratio close to unity.

\item {\bf TYC 8927-3620-1} %%49 
See \cite{2015AA...573A.126D}.

\item {\bf HIP 45336 =  $\theta$ Hya = HD 79469}   %%50
The B9.5 star $\theta$ Hya was shown to have a WD companion 
with temperature 25000-31000~K from
the analysis of the UV spectrum of the system \citep{1999AA...341..795B}.
\cite{1998ApJ...502..763V} detected low amplitude RV variations
and astrometric acceleration was detected from Hipparcos data and
from the difference of Hipparcos and historical proper motion
\citep{2005AJ....129.2420M}. Therefore, the period is expected to be or the
order of a decade, but no orbital solution is available in the literature.
We adopt the stellar masses from \cite{2013MNRAS.435.2077H}.

\item {\bf 1RXSJ091744.5+461229AB} %%51
See \cite{2015ApJS..216....7B}. Individual masses from \cite{2012ApJ...754...44J}

\item {\bf HIP 47133 = PYC J09362+3731 = GJ 9303}  %%52
Short-period SB2, see \cite{2015ApJS..216....7B} for details and references.
As for other suspected tidally locked binaries we adopt an age of 4 Gyr.

\item {\bf HIP 49669 = Regulus = $\alpha$ Leo = HD 87901} %%53
The presence of a spectroscopic companion was identified by \cite{2008ApJ...682L.117G}, with
indication that the companion is a white dwarf.
If this is the case, significant interaction between the components were expected to have 
happened, possibly explaining the extreme rotation of the (current) primary.
\cite{2009ApJ...698..666R}  modelled the evolution of the system, finding as the most likely
initial configuration two stars of 2.3 and 1.7 $M_{\odot}$ in short period (1-15 days). The current
companion to the 3.4$M_{\odot}$ component is expected to be a 0.30 $M_{\odot}$ He WD.
This scenario requires an age of the system older than 900 Myr.

The system is quadruple, as there is a close pair of low mass stars (K2V + M4V) 
at a projected separation of 175 arcsec = 4000 au, whose physical association
has been recently confirmed by \cite{2015AJ....149....8T}.
Therefore, we rely on the age indicators of the late-type component.
The lack of lithium \citep{1992AA...261..245P} indicate an age older than 500 Myr
while the chromospheric and coronal emission yield an age slightly younger than the Hyades. 
We adopt an age of 600 Myr. This estimate indicates that some adjustments are needed
in the description of the evolution of the system by \cite{2009ApJ...698..666R}, which is
not unexpected considering the theoretical uncertainties in the common envelope evolution.

\item {\bf HIP 49809 = HD 88215 = HR 3991}   %%54
This is a rapidly rotating early F star and 
single-lined SB. 

The minimum mass of the companion is $0.20~M_{\odot}$.
Stellar age is obtained through isochrone fitting.
Kinematics is compatible with young disk without association to any known group.
The star hosts a debris disk.

\item {\bf HIP 50156 = DK Leo = GJ 2079}.  %%55
The star was suspected to have RV variations in the literature 
but without conclusive evidence of binarity.
The star is also a $\Delta \mu$ binary.
We retrieved 7 spectra from SOPHIE archive, which show RV variations of about 18 km/s (peak-to-valley) 
over about 1 month. The CCF indicates a single-lined SB.
From the  small variations of RVs taken in consecutive nights 
(which is also consistent with
\cite{2010AA...514A..97L} measurements), it results that the period
is likely of the order of months.
Therefore, the large activity and fast rotation can not explained by tidal locking
but is rather due to youth.
The star was classified as a member of Columba MG and $\beta$ Pic MG 
according to \cite{2014ApJ...786....1B} and \cite{2012AJ....144..109S,2013ApJ...762...88M}, respectively.
However, the unknown system velocity represents a major source of uncertainty
in these evaluations.
Independently on the kinematics, we estimate as age of 150 Myr,
taking the lithium non-detection \citep{2010AA...514A..97L} into account.
We also adopt as tentative binary parameters to estimate the limits
of dynamical stability a period of 100d and RV semiaplitude of 10 km/s.

\item {\bf TWA 22} %%56
Originally proposed as TWA member, 
there are no adequate kinematic data according to \cite{2008arXiv0808.3362T}.
\cite{2009AA...503..281T} derived  system parallax, proposing association with the $\beta$ Pic
MG, that we adopt here. Orbit from \cite{2009AA...506..799B}.

\item {\bf TYC 7188-0575-1} %%57
See \cite{2015AA...573A.126D}.

\item {\bf CHXR 74} %%58
Binary and stellar parameters from \cite{2012AA...537A..13J}

\item {\bf TWA 5~Aab}  %%59
The central pair was first resolved  by \cite{2001ASPC..244..309M} and its orbit was 
derived by \cite{2007AJ....133.2008K} and recently refined by \cite{2013AA...558A..80K}, 
obtaining a period of 6.025 yr, a semimajor axis of 63.7 mas   and an eccentricity of 0.755.
Adopting the recently derived trigonometric parallax \citep{2013ApJ...762..118W},
the sum of the masses of the components is $0.90 M_{\odot}$ and the semimajor axis 3.2 au.
\cite{2003AJ....125..825T} identified  TWA5 as a very short period single-lined SB, with period 1.37 days and
RV semiamplitude 20 km/s. Therefore the system should include three stellar components, but some concerns on
on the existence of the short-period companion were presented by \cite{2013ApJ...762..118W}. 
An additional companion of substellar mass (TWA5 B) to the pair was 
discovered by \cite{1999ApJ...512L..69L,1999ApJ...512L..63W} 
at a projected separation of 1.95'' = 97.7 au from TWA5Aab.
The mass of TWA5B is of $20 M_{J}$ according to  \cite{1999ApJ...512L..69L} and \cite{1999ApJ...512L..63W}
and $25 M_{J}$ according to \cite{2010AA...516A.112N,2010AA...509A..52C}.
A preliminary orbital solution indicates a semimajor axis of 127 au with eccentricity of 0.24 \citep{2013AA...558A..80K}.

\item {\bf HD 102982}   %%60
Very active star, probable SB2 according to \cite{1998AJ....116..396S}.
A FEROS spectrum from ESO archive confirms the SB2 nature of the system.
\cite{2012AcA....62...67K} classified the star as a contact eclipsing 
binary with period of 0.277d.
\cite{2004AA...418..989N} gives $RV=-67.3\pm4.6$ km/s (1 measurement), which
would imply kinematic parameters typical of an old star. However, the binarity
may have significant impact on the RV.
In any case, there is a good chance that the large activity is due to
tidally-enhanced rotation and not to young age. We then adopt an age of 4 Gyr.
L05 identified an additional companion at 0.9'' (spectral type M5V), making the system triple.

\item {\bf TWA 23}  %%61
Member of TW Hya association.
RV variability was discovered by \cite{2012ApJ...749...16B}.
Their 14 measurements does not allow a unique orbital solution; they list three
equally good orbits. Conservatively, we derive the limit for dynamical stability
adopting their solution with the longest period.
We adopt the trigonometric parallax and stellar mass from \cite{2013ApJ...762..118W}.

\item {\bf HIP 59960 = HD 106906}  %%62 
Member of LCC, the star was shown to host a 11 $M_{J}$ companion at a projected separation of
650 au \citep{2014ApJ...780L...4B}. Images from JL13 were used in the discovery paper.
Very recently, Lagrange et al. 2015, A\&A, submitted, showed that the central star is an SB2 system.
The star has also a significant infrared excess, indicating the presence of a massive debris disk,
which have been recently spatially resolved with SPHERE and GPI \citep{2016A&A...586L...8L,2015ApJ...814...32K}

\item {\bf G 13-33} %%63
Young M dwarf resolved into a tight binary by B15.
The system is not associated with known moving groups.
B15 adopt an age between 10 to 300 Myr from Shkolnik et al. in prep.
We adopt 150 Myr.

\item {\bf HIP 60553}  %%64
Identified as SB2 in \cite{2006AA...460..695T}, with an estimated magnitude difference of 0.5 mag in V.
The star is also flagged as stochastic solution in the original Hipparcos catalog.
Orbital solution is not known.
Therefore, we are not able to determine whether the very large coronal emission ($\log L_{X}/L_{bol}=-2.93$) 
and fast rotation (period 0.89 days, \cite{2002MNRAS.331...45K}) are due to youth or tidal locking. 
From the lack of lithium \citep{2006AA...460..695T}, a lower limit of 400 Myr on stellar age is derived.
The space velocities derived using the single-epoch RV from \cite{2006AA...460..695T} are far from
locus typical of young stars, so we argue it is a old star tidally locked by a close companion.
We adopt and age of 4 Gyr. % Masses from mass-luminosity relation (V)

\item {\bf GJ 3729} %%65
Young M dwarf resolved into a tight binary by B15.
The system is a possible member of Tuc-Hor MG \citep{2012ApJ...758...56S}.

\item {\bf TWA 20} %%66
Young star classified as SB2 by \cite{2006ApJ...648.1206J} and \cite{2014AA...568A..26E}.
The large RV difference between the components (at least 125 km/s) indicate a
rather short orbital period.
We adopt the trigonometric parallax by \cite{2013ApJ...762..118W}.
\cite{2014AA...568A..26E} reject membership on TWA on the basis of their revised system RV.
The lithium content \citep{2009AA...508..833D} indicates an age intermediate between TWA and beta Pic MGs.

\item {\bf  HIP 62983 = HD 112131 }   %%67
Close stellar companion resolved by lunar occultation and speckle 
interferometry \citep{1975AJ.....80..689A,1996AJ....112.2260M}
The CHARM2  catalogue \citep{2005AA...431..773R} quote a projected 
separation of 0.320 arcsec %%%{\bf ???}
and brightness ratio of 5.2 in V band. From this, we infer that the
secondary is a late F star.
V12 quote a separation of 0.04  arcsec.
We adopt the projected separation from occultation for the derivation of
the critical semimajor axis for dynamical stability.
We adopt the age of 125 Myr from V12.

\item {\bf HIP 63742 = HD 113449 = PX Vir}  %%68
Member of AB Dor MG according to \cite{2004ApJ...613L..65Z} and  \cite{2008arXiv0808.3362T}.
Close companion detected by Hipparcos astrometry (with orbital solution), radial velocity
\citep{2010Obs...130..125G} and direct imaging \citep{2012ApJ...744..120E}
The orbital solution by \cite{2012ApJ...744..120E} is adopted here.

\item {\bf HIP 63962 = HD 113706}  %%69
G0 star classified as member of LCC.
J13b noted the elongated PSF, indicating an unresolved binary with projected separation well below
50 mas. The binarity is further supported by the difference among the two RV measurements available in the 
literature ($12.6\pm0.2$ km/s and $4.8\pm1.3$ km/s from \cite{2011ApJ...738..122C} and 
\cite{2007yCat..90830821B} respectively).
We tentatively adopt a projected separation of 30 mas and a mass ratio close to unity.

\item {\bf 2MASS J13215631-1052098} %%70
Close visual binary (L15), probable member of TWA following \citet{2006AJ....132..866R}.

\item {\bf HIP 66001 = HD 117524} %%71
G8 star classified as member of LCC.
J13b noted the elongated PSF, indicating an unresolved binary with projected separation below
50 mas. The binarity is further supported by the astrometric acceleration detected by Hipparcos
and the difference among the two RV measurements available in the literature (7.4 km/s and
$2.7\pm1.2$ km/s from SACY and \cite{2007yCat..90830821B} respectively).
We tentatively adopt a projected separation of 30 mas and a mass ratio close to unity.

\item {\bf HIP 72399 = HD 130260A }  %%72
See \cite{2015AA...573A.126D}

\item {\bf HIP 74045 = HD 135363 = IU Dra}   %%73
Close companion detected by B07 and GDPS ($\Delta H=4.0$).
\cite{2001AeA...379..976M} classified the star as a member of the IC2391 MG.
\cite{2007ApJ...668L.155M} support this association by noting a possible close
encounter with IC 2391 24 Myr ago with small relative velocity.
\cite{2007AJ....134.2328B} estimated an age of $35^{+14}_{-6}$ Myr from
isochrone fitting, further supporting the association.
The activity indicators and lithium abundance suggest an age
similar or younger than the Pleiades.
We adopt an age of 50 Myr.

\item {\bf HIP 76267 = $\alpha$ CrB = HD 139006} %%74
Double-lined spectroscopic and eclipsing binary composed by a B9.5 primary and G secondary.
We adopt the individual masses and orbit from \cite{1986AJ.....91.1428T}.
The space velocities are compatible with UMa membership, as previously proposed by \cite{2003AJ....125.1980K}.  %check also fuhrmann papers
The X-ray luminosity is comparable to Hyades star, if one assumes it is originating from the solar-type secondary.
We then adopt an age of 500 Myr.
A circumbinary debris disk was resolved by \cite{2013MNRAS.428.1263B} on the basis of Herschel data.

\item {\bf 1RXS J153557.0-232417 = GSC 06764-01305}  %%75
Close visual binary discovered by \cite{2008ApJ...679..762K}.
Masses of the components from \cite{2008ApJ...679..762K}.

\item {\bf HIP 76629 = HD 139084 = V343 Nor} %%76
See \citet{thalmann14}. 

\item {\bf HIP 77858}  %%77
SB1, orbit from \cite{1987ApJS...64..487L}.

\item {\bf HIP 78104 } %%78
SB1, orbit from \cite{1987ApJS...64..487L}.

\item {\bf RX J155734.4-232112 = V1148 Sco = ScoPMS 17}  %%79
The star was resolved as close visual binary in \cite{2008ApJ...679..762K} and L14,
with some discrepancy in the mass ratio between the two sources.
We adopt the individual masses by \cite{2008ApJ...679..762K}.

\item {\bf HIP 78168}  %%80
SB1, orbit from \cite{1987ApJS...64..487L}.

\item {\bf HIP 78196} %%81
A very low mass star at small separation was discovered by \cite{2015ApJ...806L...9H}
using the sparse aperture masking technique.

\item {\bf HIP 78207}  %%82
SB2 discovered by \cite{2012ApJ...745...56D}. Only single-epoch RV difference between the components
available.

\item {\bf HIP 78265 = HD 143018 }  %%83
Double-lined spectroscopic and eclipsing binary. Orbital parameters from \cite{1996Obs...116..387S}.

\item {\bf 1RXS J160210.1-2241.28 = V1154 Sco}  %%84
Short-period spectroscopic binary discovered by \cite{1989AJ.....98..987M}.
An additional system of lines at constant RV is also reported. % by  Mathieu et al. 1989.
A visual companion was discovered at about 0.30 arcsec making the system triple 
\citep{1993AJ....106.2005G,2000AA...356..541K}.
The visual component is likely the responsible for the additional spectral signature.
We adopt the stellar masses of the visual components by \cite{2008ApJ...679..762K} and the minimum mass
from the spectroscopic orbit for the unseen spectroscopic component.
The system configuration leaves little dynamical room in our planet-search zone, as
the critical semimajor axis due to the wide component is at about 12 au, corresponding to just 0.09 arcsec
at the distance of Upper Scorpius region,
while the limit for stability of planet around the whole triple system is too wide for 
being considered in this work (193 au).

\item {\bf [PGZ2001]J160341.8-200557}   %%85
SB2 discovered by \cite{2012ApJ...745...56D}. Only single-epoch RV difference between the components
available.

\item {\bf 1RXJ 160355.8-203138}  %%86
Close visual binary in Upper Scorpius.

\item {\bf 1RXS J160446.5-193031 = V1156 Sco = ScoPMS027}   %%87
Close visual binary discovered by \cite{2008ApJ...679..762K}.
Masses of the components from \cite{2008ApJ...679..762K}.

\item {\bf [PGZ2001]J160545.4-202308}   %%88
SB2 discovered by \cite{2012ApJ...745...56D}. Only single-epoch RV difference between the components
available. \cite{2012ApJ...745...56D} also reported a visual companion candidate
identified on the HIRES guide camera images, but separation and magnitude difference are not listed.

\item {\bf HIP 78977 = HD 144548 = EPIC-204506777} %%89
This is a triple eclipsing system member of Upper Scorpius association.
A close eclipsing system was originally identified by \citet{2012AcA....62...67K}. \citet{2015A&A...584L...8A} revised the period
of the short-period eclipsing binary and identified additional eclipses with a period of 33 days, 
thanks to the Kepler-2 photometric time series.
We adopt the system parameters from this latter study.
The direct imaging observations allow to probe the presence of substellar companions around the
three components of this tight triple system.
The system was also reported to have IR excess at 24 $\mu$m \citep{2011ApJ...738..122C}.

\item {\bf 1RXS J160814.2-190845 = TYC 6209-735-1 = GSC 06209-00735}   %%90
Spectroscopic binary discovered by \cite{2007AA...467.1147G}. %(P=2045 days; K=2.87 km/s e=0.20) 
The companion has also been resolved by \cite{2008ApJ...679..762K} from sparse aperture mask observations
at a projected separation 25 mas =3.6 au.
The mass of the secondary estimated by \cite{2008ApJ...679..762K} is similar to the minimum mass
derived by the RV orbital solution. % \citep[assuming the primary mass from][]{2008ApJ...679..762K}.

\item {\bf HIP 79097 = HD 144823} %%91
J13b noted the elongated PSF, indicating an unresolved binary with projected 
separation below 50 mas. At odds to HIP 63962 and HIP 66001, which were also proposed
as binaries by J13b due to PSF elongation, 
there are no multiple RV measurements available in the literature
to confirm the binarity. % {\it keep in the sample ????}
We tentatively adopt a projected separation of 30 mas and a mass ratio close to unity.
The star has an additional component at 0.8 arcsec, making the system a likely triple (J13b).

\item {\bf HIP 79404}  %%92
SB1, orbit from \cite{1987ApJS...64..487L}. Member of US.

\item {\bf 1RXS J161318.0-221251 = TYC 6213-0306-1 = BD-21 4301}   %%93
SB2 with nearly identical components discovered by \cite{2007AA...467.1147G}
These authors also derived the orbital solution. We adopt the primary mass by L14.

\item {\bf HIP 79643~B}  %%94
Triple system, formed by a F2 star, separated by 1.24 arcsec from a close pair (projected separation 47 mas, 
see L14) which is the target considered in our study. Masses from L14. Member of US

\item {\bf HIP 81266 = $\tau$ Sco = HD 149438 = HR 6165}   %%95
This early B star, member of US, was recently resolved in a close binary (projected separation 21.52$\pm$0.27 mas) 
by interferometric observations \citep{2013MNRAS.436.1694R}.

\item {\bf HIP 84586 = HD 155555}  %%96
Triple system, member of $\beta$ Pic MG. The spectroscopic binary with a period of 
1.68 days and a mass-ratio close to unity 
($M_A=1.06 M_{\odot}$ and $M_C= 0.98 M_{\odot}$) 
has a distant ($\rho=33''$) companion with $M_B=0.25 M_{\odot}$.

\item {\bf HIP 84642 = HD 155915 = V857 Ara} %97
Close binary star, possible member of Tuc-Hor association
according to \citet{2011ApJ...732...61Z} and further confirmed as 
member by \cite{2013ApJ...762...88M}
The age indicators are fully consistent with the membership assignment.

\item {\bf HIP 86346 = HD 160934}   %%98
Member of AB Dor MG.
A close companion was identified by both RV and direct imaging 
\citep{2006Ap&SS.304...59G,2007ApJ...670.1367L, 2007AA...463..707H,2010Ap&SS.330...47G,2012ApJ...744..120E}
The composite orbital solution by \cite{2012ApJ...744..120E}  was adopted.
\cite{1991AJ....101.1882W} reported a companion at 20'', confirmed by 2MASS observations \citep[see][]{2005AJ....130.1845L}. 

\item {\bf HIP 88848 = HD 166181 = V815 Her}   %%99
Triple system.
This short period spectroscopic binary (p=1.8 days) has been found by \cite{2005AJ....129.1001F} 
to have a further companion with p=5.7yr on a quite eccentric orbit (e=0.76). 
\cite{2005AJ....129.1001F} also reports a mass of $0.37 M_{\odot}$ and $0.79 M_{\odot}$ for the close and 
the distant companion respectively.
An astrometric solution is also reported, with a=4.1 au = 0.13''.
The outer companion was resolved in GDPS.
As the very high coronal activity should be induced by the close companion, we do not use
the X-ray luminosity for the age determination.
The lithium EW suggests an age of 125 Myr (to be taken with caution because of
the blending of three objects, dedicated modelling would be needed).

\item {\bf CD -64 1208 A =  TYC 9077-2489-1}   %%100
Close visual binary resolved by \cite{2007ApJS..173..143B,2010AA...509A..52C} at a projected separation of about 
0.17 arcsec and with $\Delta K=2.3$ mag.
The pair has a wide companion, the A7V star  HIP 92024=HD 172555=HR 7012 at 70 arcsec = 2000 au projected
separation, from which we took the trigonometric parallax of the system.
The system is a member of $\beta$ Pic MG.

\item {\bf HIP 92919 = HD 175742 = V775 Her} %%101
Single-lined SB (period 2.879 days, circular orbit).
\cite{2009ApJ...698.1068P} discovered $24\mu m$ excess.
The star is a BY Dra variable, with photometric period similar to the orbital one, indicating tidal locking.
This is likely responsible for the enhanced activity level of the star.
The kinematic parameters (U,V,W=24.5, 0.0, -22.6 km/s, using center of mass velocity from SB9 orbit) 
put the system far from the region of very young stars and close to UMa group.
Membership to UMa is assigned by \cite{2001MNRAS.328...45M} and considered possible by \cite{2003AJ....125.1980K}
J13 adopted an age of 40-60 Myr from \cite{2009ApJ...698.1068P}. 
Marginal detection of lithium have been reported by \cite{2000AAS..142..275S} and 
\cite{2007AJ....133.2524W} %(29 and 20 mA respectively)
while only upper limits by \cite{2012AA...547A.106M}.
These values are compatible with a star of the age of UMa.
We then adopt 500 Myr.

\item {\bf HIP 94050 = HD 177996:} %%102
The star is a short-period SB2 \citep{1998AJ....116..396S},
but the orbital solution is not available. The line depth ratio is about 0.5 at 6700 \AA.
Lithium was detected, with an EW likely larger than Hyades of
similar color, indicating a true moderately young star rather than a tidally-locked system.  
Adopting a RV of -38.4 from \cite{1998AJ....116..396S},
kinematic parameters similar to the Hyades are derived (UVW = -40.5, -14.2, 5.3).
We adopt an age of $400\pm200$ Myr.

\item {\bf HIP 94863  = HD 180445}  %%103
The star is a short period SB2 \citep[G8V+K5V]{2002AA...384..491C}.
A preliminary orbital solution was provided by \cite{2006AA...450..681T}.
They also found evidence for a wide-separation tertiary component
at 9.4 arcsec. 
%{\it check in L05 images}. 
We adopt the individual masses from MSC
and we assume circular orbit due to the short period.
No lithium was detected by \cite{2002AA...384..491C, 1998AJ....116..396S}, with limits 
corresponding to ages older than about 500 Myr.
The rotation period by \cite{2012AcA....62...67K} %2.455 
is very close to the orbital period.
Therefore the large coronal and chromospheric emission appear to be due
to tidally induced rotation and not to young age.
The main sequence status of both components and the thin disk kinematics
put an upper limit of about 8 Gyr. We then adopt an age of 4 Gyr.
% no data in ESO archive

\item {\bf HIP 95149 = HD 181321 = GJ 755}   %%104
Reported as a SB in \cite{2004AA...418..989N} (scatter of RV of 2.3 km/s over 
about 9 years) and \cite{2007astro.ph..1293G} (trend of -1.4 km/s/yr, with a possible 
curvature over 1.2 yr). The astrometric acceleration was also detected by Hipparcos.
We then argue that the companion is most likely a low mass star with a period
of several years.
 A spectral type later than K5 is expected from the lack of
signature of the secondary in the spectra \citep{2002AA...384..491C}.
The age indicators point to a moderately old age, compatible with
membership in Castor MG proposed by \cite{2003AA...400..297R}.

\item {\bf HIP 97255=HD 186704}   %%105
This star shows RV variations of at least 5 km/s peak-to-valley
\citep{2002ApJS..141..503N,2004AA...418..989N,2010CoSka..40...83T}.
\cite{2014AJ....147...86T} quote a period of  3990d from a priv. common by D. Latham.
As the minimum mass of the spectroscopic companion is not included in this study,
we adopt $0.3 M_{\odot}$ for the calculation of the stability limits
The star has a wide companion, the flare star V1406 Aql,  at 9 arcsec.
\cite{2013ApJ...778....5Z} classified the system as a probable member of the Octans-Near
Association. The age indicators are compatible with an age similar to the Pleiades.

\item {\bf 2MASSJ19560294-3207186}  %%106
Close visual pair with an additional component, the M0 star TYC 7443-1102-1, at 26 arcsec.
This triple system is a probable member of $\beta$ Pic MG.

\item {\bf HIP 100751 = HD 193924 = $\alpha$ Pav }  %%107
This star, member of Tucana association, is a close spectroscopic binary (SB9). 
The minimum mass of the companion is $0.26~M_{\odot}$.

\item {\bf HIP 101800 =  $\iota$ Del = HD 196544}   %%108
This star is a spectroscopic binary with an Am primary.
The short-period orbit from SB (P=11.039d; e=0.23) yields a 
minimum mass of $0.49~M_{\odot}$  for a primary stellar mass of 
$2.0~M_{\odot}$. The companion is then most likely a early M or a K dwarf.
The star also shows IR excess \citep{2007ApJ...660.1556R,2011ApJ...730L..29M}.
This star was observed in deep imaging by R13 and B13. These studies 
provide discrepant age values. R13 assumed an age of 30 Myr from \cite{2007ApJ...660.1556R} while 
N13 list their own determination of a median age of 272 Myr (69-444 Myr 95\% limits).
To further investigate the issue, we consider the kinematic of the system,
adopting the center of mass velocity from SB9 and distance and proper motions from
\cite{2007AA...474..653V}. the space velocities results $U,V,W= -7.7$, $-4.2$,  $-8.2$.
These are quite far from those of the moving groups younger than 100 Myr
and compatible within error with those of the Castor MG \citep{2003AA...400..297R}.
An additional indirect evidence against a very young age comes from the lack of 
detection of the system (whose X-ray emission should be dominated by the secondary, unless it is a WD) using ROSAT. 
Therefore, we adopt the age of the Castor MG as given in \cite{2003AA...400..297R} (320 Myr).

\item {\bf TYC 5206-0915-1}  %%109
See \cite{2015AA...573A.126D}.

\item {\bf HIP 105404 = HD 202917 = BS Ind}   %%110
See \cite{thalmann14}.

\item {\bf  HIP 105441 = HD 202746 = V390 Pav}  %%111
This star was classified as a new potential member of Tuc-Hor association in
\cite{2001ApJ...559..388Z} but it was rejected by \cite{2003ApJ...599..342S} because
of its low lithium content.
Membership to $\beta$ Pic MG is instead supported by \cite{2013ApJ...762...88M}.
Radial velocity is variable with peak-to-valley difference of at least 30 km/s
\citep{2004AA...418..989N,2006AA...460..695T,2006AJ....132..161G} and
\cite{2006AJ....132..161G} noted the possible presence of blending in the violet
part of their optical spectrum.
The star has a wide (26 arcsec projected separation) companion, 
TYC 9114-1267-1. Both components were observed by K07.
The very similar proper motions and the fact that the photometric distance   of 
TYC 9114-1267-1 is compatible with the trigonometric parallax of HIP 105441 
suggest physical association, with the
RV difference being due to binarity of the primary.
However, TYC 9114-1267-1 (K7V) has detectable lithium (EW=15 mA) in SACY, suggesting
an age of about 30-50 Myr, while HIP 105441 (K2V) has no detectable lithium
\citep{2003ApJ...599..342S,2006AA...460..695T}, corresponding to a lower limit 
on stellar age of about 400 Myr.
HIP 105441 shows indication of enhanced activity \citep{2011ApJ...734...70A} and rotation \citep{2012AcA....62...67K}, 
which would indicate age of about 100 Myr, but considering the
lack of lithium, we favour tidal locking as the source of these characteristics.
We then adopt an age of 4 Gyr but we note that further studies are needed for
a characterisation of this object and to investigate its physical association with
TYC 9114-1267-1.

\item {\bf HIP 107556 =  $\delta$ Cap = HD 207098 = GJ 837}  %%112
See \cite{thalmann14}.

\item {\bf [FS2003] 1136 = 1RXS J214906.4-641300}  %%113
Resolved as a close visual binary by CH10.
Stellar parameters from CH10.

\item {\bf HIP 108195:} %%114
Triple system formed by a close pair of F stars and a M5-M7 
companion at 4.9 arcsec, identified by \cite{2010AA...509A..52C}. 
The inner pair has a preliminary orbit in WDS.
The system is a member of Tucana association.

\item {\bf HIP 109901 =  HD 211087 =  CS Gru}  %%115
See \cite{thalmann14}.

\item {\bf GJ 860 = HD 239960~B } %%116
Close visual binary with orbital solution.
Individual masses from \citet{1999ApJ...512..864H}. Age from H10.

\item {\bf PPM 366328 = TYC 9129-1361-1}   %%117
Classified as possible member of Tuc-Hor association \citep{2001ApJ...559..388Z},
\cite{2006AA...460..695T} showed that instead the star is an SB2 ($\Delta V=1$ mag) with
no detectable lithium, indicating an age older than the Hyades.
The different RVs as measured by \cite{2000ApJ...535..959Z} and \cite{2006AA...460..695T} also support binary.
The very fast rotation ($v \sin i = 88$ km/s) and bright X-ray emission ($\log L_{X}/L_{bol}=-3.38$)
are then likely due to tides of the companion.
There is another companion at 24 arcsec \citep{2003AN....324..535N}, classified as M2 by \cite{2006AJ....132..866R}. 
Taking into account both components, we adopt a distance of 60 pc, and, assuming tidal locking as responsible of the
enhanced activity of the SB2 system, an age of 4 Gyr. Masses of the components from mass-luminosity relations.

\item {\bf HIP 116003 = GJ 1284 = 2MASSJ23301341-2023271}  %%118
This star was classified as SB2 by \citet{2006AA...460..695T}. The orbital solution is
not available.
It was classified as candidate member candidate of Columba association by \citet{2014ApJ...788...81M}
We adopt the trigonometric distance by \citet{2014AJ....147...85R}.

\end{enumerate}

\end{document}

%% file: mastertable.tex
\setcounter{table}{1}

\small{
\tiny{
\begin{landscape}
\label{tab:master}
\begin{longtable}{lllr|rrrlr|rrrrr|l}
\caption{CBIN sample } \\
\hline\hline % inserts double horizontal lines
\# &  Star ID            &  RA(2000)   &  Dec(2000)   & Dist  &  Age   &    H    &   SpT  & $M_A$     &  $M_B$  &  $\rho$     &  p     & ecc  & $a_{crit}$  & Ref   \\
    &             &             &              &  (pc) & (Myrs) &  (mag)  &        & ($M_{\odot}$)  & ($M_{\odot}$) &  (arcsecs) & (days) &      &  (AU)     &      \\   
\hline
\endfirsthead
\caption*{Table 2, continued.}\\
\hline\hline
\# & Star ID            &  RA(2000)   &  Dec(2000)   & Dist  &  Age   &    H    &   SpT  & $M_A$        &    $M_B$   &  $\rho$     &  p     & ecc  & $a_{crit}$ & Ref   \\
    &            &             &              &  (pc) & (Myrs) &  (mag)  &        & ($M_{\odot}$)  &  ($M_{\odot}$) &  (arcsecs) & (days) &      &  (AU)     &       \\   
\hline
\endhead

\hline
\endfoot

\hline 
\caption*{%Properties of the targets. See App.~\ref{app:notes} for notes on individual objects and references.
%If flagged with $^*$, See App.~\ref{app:notes} for individual notes. \\
%Distance from \cite{2007AA...474..653V}, unless flagged with t: \citep[distance from][]{2008arXiv0808.3362T} or  s (distance from the source paper). \\
%Age from source paper, unless flagged with a: this paper (see App.~\ref{app:notes});  b: \cite{2008arXiv0808.3362T}; \textbf{ c}: \cite{2003AeA...400..297R};  d: \cite{2006ApJ...649..894L}. \\
%The masses are the ones adopted in the source paper listed in Tab.~\ref{tab:sample}, unless indicated by one 
%of the following flags: 
%\textbf{a:} $M_{star}$ from individual papers (see App.~\ref{app:notes}); 
%\textbf{b:} $M_{star}$ from \cite{1997AJ....113.2246R,2000A&A...364..217D};
%\textbf{c} $M_{star}$ from \cite{2007A&A...475..519H};
%\textbf{d} $M_{star}$ assumed; 
%\textbf {h:} the star is part of a hyerarchical system, and the listed mass is the sum of 
%the masses of the close pair, reported in App.~\ref{app:notes}.
%The \textbf{w} flag indicates the presence of a wide companion, as listed in Tab.~\ref{tab:widebin}.
}
\endlastfoot
% Name                     &  RA(2000)   &  Dec(2000)      & Dist         &  Age       &    H    &   SpT  & $M_{star}$     &  $M_{comp}$  &  $\rho$    &  p     & ecc  & $a_{crit}$ & Note & Ref   \\
 1 & TYC 5839-0596-1            & 00:12:07.60 &  -15:50:33.00   &  33.4        & 4000       & 6.64   & K0IVe     & 0.70     &  0.70        &   --       &    --  &  --  &   $<10$    & NLP \\
 2 & HIP 3210                   & 00:40:51.58 &  -53:12:35.70   &  44.80       &  150       & 6.22   & F7V       & 1.14     &  0.40        &   --       &    --  &  --  &   $<10$    & SONG \\ 
 3 & HIP 3924                   & 00:50:24.31 &  -64:04:04.02   &  53.20       &  500       & 6.71   & F7V       & 1.17     &  0.80         &   --       &    --  &  --  &   $<10$   & NLP \\ 
 4 & HIP 4448                  & 00:56:55.47 &  -51:52:31.86   & 40.63    &   50       & 6.52   &  K3/K4V   &  0.79     &  0.75     &   0.228    &   --     &  --      &   34.0      & K07  \\ % updated
 5 & 2MASS J01033563-5515561    & 01:03:35.6  &  -55:15:56.1    &  47.2          &   45       &  9.58    &  M7      &   0.19     &     0.17       &    0.25       &    --       &    --     &      43.1      & L15  \\ %% NEW justine
 6 & HIP 4967             & 01:03:40.12 &  +40:51:29.25   &  29.9   &    149   &    7.46    &    M0       &  0.65    &     0.23      &   0.27     &    --     &  --     &    31.0        & B15 \\
 7 & HIP  9141                  & 01:57:48.90 &  -21:54:05.00   & 40.90    &   45       & 5.56   &  G4V     &  0.91 &  0.86     &   0.15     &    --     &  --     &  22.5      & B07,J13, SONG \\ % updated 
 8 & NLTT 6549            & 01:58:13.61 &  +48:44:19.7    & 44.0     &  625       & 8.24       &  M1.5    &  0.60     &   0.34        &     0.050  &    --     &   --    &     8.4        & B15 \\ 
 9 & HIP 11072                  & 02:22:32.55 &  -23:48:58.80   & 21.96    & 5800       & 3.71	&  G2V     &  1.20  &  0.96      &   0.54    &  25.81yr  &  0.339  &  38.4     & GDPS \\ 
10 & HIP 12413            & 02:39:47.99 &  -42:53:30.03   &  39.8        &   42   &  4.62      & A1V     &   2.03     &    $\sim$0.6   &   --       &   --   &  --  &  $\sim$20    &  V12 \\ 
11 &HIP 12545                  & 02:41:25.90 &  +05:59:18.40   &  42.03       &   24   & 7.20   & M...      & 0.76 &    --         &   --       &    --  &   -- &   $<20$   & B07,B13, SONG   \\
12 & HIP 12638                  & 02:42:21.31 &  +38:37:07.23   &  45.45   &  149      & 7.10   &  G2V     &  0.91  & 0.28       & 
 0.026     & 466.5d  & 0.084 &    3.0    &  B13 \\ 
13 & HIP 13081                 & 02:48:09.14 &   +27:04:07.10   & 24.63    &  500       & 5.69	&  K1V     &  0.98  &  0.18      &    --     &   17 yr  &  0.69    &  26.3     & GDPS  \\ 
14 & HIP 14555                  & 03:07:55.75 &   -28:13:10.97   &   19.2   &  70      &    6.58     &   K8V   &  0.6   &     0.6  &      --     &  --      &   --     &  $<1$     & M14 \\   %%%%%%%%% NEW
15 & HIP 14576                  & 03:08:10.13 &   +40:57:20.30   &  28.5    &  460       & 1.95   &  B8V      & 4.51  & 1.70     &   --       &  679.9  & 0.22 &   8.67   & JB11  \\
16 & HIP 16247                  & 03:29:22.88 &  -24:06:03.10   &  31.06       & 4000   & 6.59   & K3V       &  0.70 & 0.64     &   --       &    3.98 &  0.00  &   0.13    & L05   \\
17 & 2MASS J03363144-2619578    & 03:36:31.4  &  -26:19:57.8      &  44          &   45       &  9.80    &  M6      &   0.18     &     0.07       &    $\le$0.12       &    --       &      --   &   $\le$ 20.3       & L15  \\ %% NEW justine acrit updated
18 & HIP 16853                  & 03:36:53.40 &  -49:57:28.90   &  41.70       &   45       & 6.26   &  G2V      &  1.0     &  0.40       &   --       &  201.0  & 0.00 &   1.77    & SONG \\
%%HD 285281                       &   04:00:31.069 & +19:35:20.85   &   140      &     {\bf 20}    &  7.75       &    K1     &  {\bf 1.11}      &   {\bf 1.1}       &    --        &  --        &  --        &  {\bf $<$ 10 }        &   D15 \\   
%V773 Tau                        &    04:14:12.922 &  +28:12:12.30  &   {\bf 140}      &    {\bf 2}          &         &         & {\bf 1.5+1.1}     & {\bf 0.7}         &    46yr        &  0.1       &  0.30        &   {\bf  66  }     &   D15 \\ TOGLIERE V773 Tau 
19 & HD 282954 & 03:46:38.77  & +24:57:34.69   & 133.5 & 125 & 8.85  & G0V    & 0.75 & -- & -- & -- & -- & $<10$ & Y13 \\
20 & HII 1348  & 03:47:18.06  & +24:23:26.80   & 133.5 & 125 & 11.02 & K5V+M8 & 0.67 & 0.55 & -- & -- & -- & $<10$ & Y13 \\ 
21 & HD 23863  & 03:49:12.18  & +23:53:12.46   & 133.5 & 125 & 7.60  & A7V    & 1.75 & 0.45 & 0.022 & -- & -- & 11.3 & Y13 \\ 
22 & HIP 19176 & 04:06:38.80  & +20:18:11.13   & 108.2 &  25 & 8.20  & F8/G1  & 1.14 & --   & --    & --  & -- & $\sim$40 & D15 \\
23 & RX J0415.8+3100            &  04:15:51.38 &+31:00:35.6 &    200  &   100      &   10.05      &     G6      &   0.95 & $>$0.21          &   --         &     --     &   --     &  $<$1        &   D15 \\ 
24 & RX J0435.9+2352     & 04:35:56.83 &   +23:52:05.0    &  140              &  20    &   8.95      &   M1.5      &  0.42       &    0.27     &  0.086      &  --      &  --  &  45.4   & D15 \\      
25 & HIP 21482                 & 04:36:48.24 &   +27:07:55.90   &  18.00$^n$   &  625       & 5.40   &  K2V      & 0.84     & 0.19        &   --       &   1.79  & 0.00 &   0.06    & GDPS  \\
26 & GJ 3305               & 04:37:36.13 &  -02:28:24.77   &  29.42       &   24   & 4.77    &  M0.5V   & 0.85  & 0.50       &  0.09  &   21.5yr & 0.06     &   22.38   &   D12, BN13, L15   \\
%%%%RXJ 0441.4+2715                  & 04:41:24.00  & +27:15:12.4     & 140.00       &  2         & 10.57    & G8     & 6.74  &   --    &        0.065    &   --       &   --       &  39.7    &   D15 \\  acrit>50
27 & HIP 21965                       & 04:43:17.20   & -23:37:42.04   &  63.6          &     45           &     6.07        & F2-3IV/V   &  1.42       &    0.63         &    --        &   709        &  0.29          &      6.6             & M15 \\
28 & DQ Tau                          & 04:46:53.063  &  +17:00:00.10  &  140      &   2       &   8.54      &   M1V      &  0.56 &  0.56     &    --        &   15.80       &   0.556    &   0.47        &   D15 \\ 
29 & HIP  23296                 & 05:00:39.80  &  -02:03:57.70   &  49.60       &  125       & 5.62   & A8IV      & 1.50     & 0.29       &   --       &   8.111  & 0.00 &   0.20    & V12   \\ 
30 & HIP 23418        & 05:01:58.79  &  +09:58:59.29  &  24.6        &   24      &  6.66      &    M3.5       &  0.26      & 0.15    &   --         &   12d      &  -- &  0.29  & L15 \\ % 
31 & L449-1AB		   & 05:17:22.93 &  -35:21:54.50   & 11.85	  &   500     & 6.85   &  M4.0	   & 0.33 & 0.24 &  0.047     &	  --	 &     --   &   2.00     & B15   \\ % updated
32 & HIP  25486                 & 05:27:04.76 &  -11:54;03.47   &  27.04       &   24  & 2.93   &  F7V      & 1.06  & 0.76   &   --       &   --     &   --  &   $<10$  & K07,L05,R13 B13,BN13\\
33 & AB DorAC               & 05:28:44.80 &  -65:26:54.90   &  15.10       &   149   & 4.80   &  K2Vk     & 0.865     &  0.09      &  0.156 &  11.76yr & 0.60  &  20.24   & B07   \\
34 & AB DorBab              & 05:28:44.30 &  -65:26:46.00   &  15.10       &   149   & 7.66   &  M3.5     & 0.17      &  0.15      &  0.06  &    --    &  1.19 &   4.01   & CH10  \\
35 & 2MASS J05320450-0305291    & 05:32:04.5  &  -03:05:29      &  42          &   16       & 7.24     &  M4      &   0.42     &       0.25     &  0.18         &      --     &    --     &    44.5       & L15  \\ %% NEW justine
36 & HIP  30920~A               & 06:29:23.40 &  -02:48:50.30   &   4.10       & 150   & 5.75   &  M4       & 0.20      & 0.10       &  1.04  &    5889d     & 0.37  &  15.09   & GDPS  \\	      
37 & HIP  32104                 & 06:42:24.33 &  +17:38:43.11   &  43.60       &   42       & 5.07   &  A2V      & 2.70      & 0.51       &   --       &    522  &  0.10 &  4.76    & B13 \\ 
38 & HIP  35564                 & 07:20 21.42 &  -52:18 41.50   &  31.70       &  250       & 5.13   &  F5       & 1.46      &  0.73      &   --       &    --   &   --  &  $<10$   & NLP \\
39 & HIP  36349                   & 07:28:51.5  &  -30:14:47      &  15.7          &   149       & 5.97     &  M1      &   0.48     &    0.25   &    0.46       &    --       &  --       &  27.7      & L15  \\ %% NEW justine % updated
40 & HIP  36414                 & 07:29:31.41 &  -38:07:21.60   &  52.50       &  250       & 6.51   & F7V       & 1.23      &  0.61      &   --       &    --   &   --   &  $<10$  & NLP \\ 
41 & GJ 278~C                     & 07:34:37.58 &  +31:52:11.05   &  14.90       &  320       & 5.42   &    --       & 0.60      & 0.60       &   --       &    0.82  &  0.00 &  0.04   & H10  \\ 
42 & HIP  38160                 & 07 49 12.90 &  -60:17:01.28   &  34.60       &  250       & 4.86   &  F1      &  1.50  &  0.65      &  0.141     &    --    &   --    &  18.8     & R13 \\  % updated
43 & HIP  39896                 & 08:08:56.41 &  +32:49:11.14   &  21.30       &   42       & 6.58   &  K7      &  0.55  &  0.45      &  0.252     &    --    &   --    &  20    & B13  \\ % updated
44 & HIP  39896~B               & 08:08:55.44 &  +32:49:05.10   &  21.30       &   42       & 7.36   & M2.8+M3.3 & 0.36      & 0.36       &   --       &    --    &  --    &  $<3$  & SONG \\  
45 & EM Cha                 & 08:43:07.24 &  -79:04:52.50   &  97.00    &   11       & 7.75    & K7Ve     & 1.0        & 0.4        &   --       &    2.6d   &  --   &  0.10  & B06   \\
46 & RS Cha                 & 08:43 12.20  &  -79 04 12.30   &  97.00    &   11       & 5.87    & A8V+A8V  & 1.89       & 1.87       &   --       &    1.67  & 0.0   &  0.10  & B06   \\
47 & EQ Cha                 & 08:47 56.77 &  -78 54 53.20   &  97.00    &   11       & 8.68    & M3.2Ve   & 0.40   & 0.40   &  0.04      &   --     & --    & 14.1  & B06   \\% updated
48 & TYC 8927-3620-1            & 08:58:48.60 &  -61 15 15.00   &  81.80       &   20      & 7.65    & G8IV     & 0.89       & 0.86       &  0.087     &   --     & --    & 26   & NLP  \\% updated  
49 & HIP  45336                & 09:14:21.86 &  +02:18:51.34   &  34.80       &  130      & 4.04    & B9.5+WD  &  2.52       & 1.21  &   --       &  --        & --    & $\sim$25 & N13,JB11   \\
50 & 1RXSJ091744.5+461229AB & 09:17:44.73 &   46:12:24.70   & 32.00	  &  50	  & 7.49	&  M2.5	   & 0.48    & 0.35	 &  0.20      &  --	 &     -   &   24.1   & B15   \\  % % updated
51 & HIP 47133		   & 09:36:15.93 &   37:31:45.70   & 33.7	  &  4000	  & 7.43	&  M0.5	   & 0.58    &  0.58	 &  --	      &	 --  	 &    -- 	   &   $ < 1$  & B15   \\ 
52 & HIP  49669                 & 10:08:22.31 &   +11:58:01.90   &  23.80    & 600   & 1.66   & B7V      & 3.40    & 0.30   &   --       &  40.11   & 0.00  &    0.67  & JB11,N13  \\
53 & HIP  49809                 & 10:10:05.89 &  -12:48:57.32   &  27.70       & 800      & 4.46   & F3V      & 1.41        & 0.2        &   --       & 28.10    & 0.07  &    0.49  & J13  \\ 
54 & HIP  50156                 & 10:14:19.18 &  +21:04:29.55   &  23.10       &   150    & 6.45   & M07V     & 0.61  & $\sim$0.19 &   --       & $\sim$100  & --    &  $\sim$1.5   & B13, B15, SONG  \\
55 & TWA 22                 & 10:17:26.90 &  -53:54:28.00   &  18.00    &  24   & 8.08   & M6       & 0.12    & 0.10   &  0.10      &  5.15yr  & 0.09  &    4.77  & CH10, SONG  \\
56 & TYC 7188-0575-1            & 10:22:04.50 &  -32:33:27.00   &  43.20       & 4000      & 7.385  & K0V:e    & 0.71        &  0.35      &   --       &    --    &   --  &  $<10$   & NLP \\
57 & CHXR 74                    & 11:06:57.33 &  -77:42:10.67   & 160.00    &   2  &10.51  & M4.25     & 0.24        & 0.08       &  0.023     &    4770d      & 0.00  &    8.88  & JJ12  \\
58 & TWA 5Aab                   & 11:32:50.26 &  -34:36:27.23   &  50.10       &   10   & 7.35  & M8.5     &  0.39       &  0.51      &  0.0637     &  6.025yr  & 0.755 &   12.49  & L05   \\
59 & HD 102982              & 11:51:09.14 &  -51:52:32.30   &  62.1    &4000   & 6.96   & G3V      & 1.09    &   1.09   &   --       &  0.277  &   --   &   0.03  & L05   \\ 
60 & TWA 23                     & 12:07:27.40  &  -32:47:00.00   &  53.90       &    10       & 8.02  & M1       & 0.35        &  0.12     &   --       & 1522     & $\sim0.7$ &  8.4 & CH10, SONG, L15 \\
61 & HIP 59960                 & 12:17:53.19 &  -55:58:31.89   &     92.1        &   17       &  F5V    & 1.37     &  1.34  &    --            &   --        &    --         &   --       & $\sim$2      &  JL13 \\  %%% HD106906 NEW 
62 & G13-33             & 12:22:50.62 &  -04:04:46.24   &  15.00       &   150     & 9.11      & M4.5     &   0.13      &    0.12   &    0.09    &   --     &   --    &    4.9      &  B15  \\
63 & HIP 60553                 & 12:24:47.30 &  -75:03:09.40   &  72.62       & 4000   & 7.80  & K3Ve     & 0.92    & 0.86  &   --       &   --     &   --    &    $<1$  &   B07   \\
64 & GJ 3729             & 12:29:02.90 &  +41:43:49.7    &  17.00  &  45  &    8.18       &   M4     &     0.26 & 0.18    &  0.050     &   --  &      --    &   3.2    & B15 \\     
65 & TWA 20                    & 12:31:38.07 &  -45:58:59.4    &   77.5  &  15  &    8.693  &   M2     &  0.50       &  0.45      &     --        &     --     &    --     &  $<2$    &  BN13    \\ %%%% NICI 
66 & HIP  62983                 & 12:54:18:70  &  -11:38:54.90   &  68.50       &  125       & 5.85  & A2V      & 2.20        & 1.29      &  0.10     &  --     &   --    &   26.2  & V12   \\ % updated
67 & HIP  63742             & 13:03:49.65 &  -05:09:42.50   &  21.69       &   149    & 5.67  & G5V     & 0.84    & 0.51   &  0.034     & 216.9   & 0.30     &    2.43  & GDPS,H10, SONG  \\ 
68 & HIP 63962                  & 13:06:27.40 &  -56:52:44.83   & 236.4       &   17       & 7.88  &  G0      & 1.36  & $\sim$1.3     & $\sim0.03$ &   --     &  --      &  $\sim26$  & JL13  \\ % updated
69 & 2MASS J13215631-1052098    & 13:21:56.3  &  -10:52:09.8      &  40          &   11       &  8.82    &  M4.5      & 0.15     &  0.075     &   $\le$0.11        &     --      &     --    &   $\le$16.8   & L15  \\ %% NEW justine  acit updated
70 & HIP 66001                 & 13:31:53.62 &  -51:13:33.20   & 152.4       &   17       & 8.01  &  G8      & 1.23       & $\sim$1.2     & $\sim0.03$ &   --     &  --      &   $\sim17$   & JL13  \\ % updated
%{\bf HIP 69562}            & 14:14:21.36 &  -15:21:21.75   &  30.2  &            &       &         &             &          &  0.32      &    --  &    --      &      & SONG  \\ 
71 & HIP  72399                 & 14:48:09.65 &  -36:47:02.00   &  46.10       &  500       & 7.485 & K3V(e)   & 0.75        & 0.37     &   --       &    --    &   --     &  $<10$   & NLP \\
72 & HIP  74045                 & 15:07:56.30 &   +76:12:02.70   & 28.79    &   50       & 6.33   &  G5      &  0.96  &  0.59      &  0.302     &    --    &   --    &  33.1     & B07; L05; GDPS \\  % updated
73 & HIP  76267                 & 15 34 41.27 &  +26:42:52.89   &  23.00       &  500       & 2.39   & B9.5IV+G & 2.58       &  0.92     &    --      & 17.36   & 0.37     &    0.71  & J13 \\ 
74 & 1RXS J153557.0-232417      & 15:35:57.80 &  -23:24:04.60    & 145.00       &   11       & 9.60   & K3      &  0.99       &   0.10     &   0.05468  &   --   & --       &   29.7  &      L14 \\ % updated
75 & HIP 76629                  & 15:38:57.54 &  -57:42:27:34    &  38.54       &   24       &  9.45    & K0V     &  1.12  & $\sim$~0.11 &   --  &  $\sim$~4.5yr & $\sim$~0.5 & $\sim$~10.9 & BN13   \\
76 & HIP 77858                  & 15:53:53.92 &  -24:31:59.20    & 128.87       &   11       &  5.38   & B5V     &  4.20 &   0.50     &    --      &  1.92   & 0.36      &   0.17     & L14 \\
%{\bf RXJ155629.3-234821}   & 15:56:29.417 & -23 48 19.75    & 145.0        &   11       &        & M1.5     &  0.54 &   0.34     &    0.092   & --      &  --       &    50.4       & L14 \\ NO acrit>50
77 & HIP 78104                  & 15:56:53.07 &  -29:12:50.80    & 144.72       &   11       &  4.52  & B2IV-V   &  7.80 &   0.48     &    --      &  4.0    & 0.27      &   0.30     & L14 \\
78 & RX J155734.4-232112        & 15:57:34.31 &  -23:21:12.30    & 145.00       &   11       &  9.23  & M1V      &  0.60 &   0.32     &   0.05385  &   --     & --        &   30      & L14 \\% updated
79 & HIP 78168                  & 15:57:40.46 &  -20:58:59.20    & 141.24       &   11       &  5.77  & B3V      &  5.90 &   2.12     &    --      &   10.0   &  0.58     &   0.73     & L14 \\ 
80 & HIP 78196                  & 15:57:59.35 &  -31:43:44.15    & 126.7        &   11       &  7.12  & A0V      &  2.46 &   0.10     & 0.074   &   --     &   --      &   34.4     & L14 \\ %%% NEW % updated
81 & HIP 78207                  & 15:58:11.36 &  -14:16:45.50    & 143.47       &   11       &  4.83  & B8Ia/Iab &  2.90 &   2.90     &     --     &   --     &    --     &   $<$5      & L14 \\ 
82 & HIP 78265                  & 15:58:51.11 &  -26:06:50.70    & 179.53       &   11       &  3.50  & B1V+B2V  & 10.0  &   6.33     &     --     &  1.57   & 0.00      &   0.16      & L14 \\ 
83 & 1RXS J160210.1-2241.28     & 16:02:10.45 &  -22:41 28.00    & 145.00       &   11       &  8.26  & K5IV     &  0.87 &   0.48     &     --     &  2.4    & 0.024     &  0.098      & L14 \\
84 & PGZ2001 J160341.8-200557   & 16:03:41.87 &  -20:05:57.80    & 145.00       &   11       &  9.76  & M2       &  0.37 &   0.37     &     --     &   --     &  --       &   $<$5     & L14 \\  
85 & 1RXJ 160355.8-203138 & 16:03:54.964 & -20:31:38.38    & 145.0        &   11       &  8.89      & M0       &  0.61 &   0.56     &    0.078   & --      &  --       &    33.5       & L14 \\ 
86 & 1RXS J160446.5-193031      & 16:04:47.76 &  -19:30:23.10    & 145.00       &   11       &  8.27  & K2IV     &  1.12 &   0.74     &   0.04318  &   --     &  --       &  23.7      & L14 \\  % updated
87 & PGZ2001 J160545.4-202308   & 16:05:45.40 &  -20:23:08.80    & 145.00       &   11       & 10.75  & M2       &  0.37 &   0.37     &     --     &   --     &  --       &   $<$5     & L14 \\  
88 & HIP 78977                  & 16:07:17.79 &  -21:55:36.30    & 116.70       &   11       &   7.15     & F8V      &  1.44 &   1.93     &     --     &  33.945  &  0.265  &   0.39    & J13  \\ 
89 & 1RXS J160814.2-190845      & 16:08:14.74 &  -19:08:32.80    & 145.00       &   11       &  8.60  & K2       &  1.12 &   0.21     &    0.0246  &  2045    &  0.20     &  10.14     & L14 \\   
90 & HIP 79097                  & 16:08:43.66 &  -25:22:36.70    & 200.8        &   11       &  7.33  & F3       &  1.56 &  $\sim$1.5      & $\sim0.03$ &   --     &  --       &   $\sim$18    & JL13 \\ % updated
%% HIP 79374                  & 16:11:59.74 &  -19:27:38.10   & 145.35       &   11       &  3.80  & B2IV     &  7.80 &   3.59     &   0.072    &   --     &  --       &  40.34     & L14 \\ % 3pla, updated NO ZONE STABILI
91 & HIP 79404                  & 16:12:18.21 &  -27:55:35.00    & 146.84       &   11       &  5.01  & B2V      &  7.80 &   1.12     &    --      &  5.78   &  0.19     &   0.37     & L14 \\ 
92 & 1RXS J161318.0-221251      & 16:13 18.59 &  -22:12:48.90    & 145.00       &   11       &  7.59  & G9       &  1.70 &   1.65     &    --      &  166.9  &  0.226    &   2.7      & L14 \\
93 & HIP 79643 B                 & 16:15:09.27 &  -23:45:34.80    & 210.97       &   11       &  8.15  &  --        &  0.76 &   0.29     &  0.047     &   --     &  --       &  38.32     & L14 \\ % updated
94 & HIP 81266                  & 16:35:52.96 &  -28:12:57.70    & 145.35       &   11       &  3.48  & B0V      & 16.0  &   4.80     &   0.02152  &   --     &  --       &  12.08     & L14 \\% updated
95 & HIP  84586             & 17:17:25.50 &  -66:57:04.00   &  31.45       &   24    & 4.91  & K1       & 1.059   & 0.986  &    --      &  1.68  & 0.00     &    0.07  & CH10, SONG  \\ 
96 & HIP 84642            & 17:18:14.65 &  -60:27:27.52   &    58.9          &    45    &      &  G8V     &   0.90   & 0.40    &  0.22  &  --     &  --       &    49.7        & SONG  \\
97 & HIP  86346                 & 17:38:39.81 &   61:14:14.00   &  33.12       &   149    & 7.00  & K7       & 0.69 & 0.54 &  0.213 & 3764    & 0.636     &   19.90    & GDPS  \\
98 & HIP  88848             & 18:08:16.03 &   29:41:28.10   &  34.38       &   125    & 5.76  & G6V       & 1.27 & 0.79 &  0.063 & 5.75 yr  & 0.765     &   17.13    & GDPS  \\
99 & CD -641208Aab          & 18:45:37.00 &  -64:51:44.60   &  28.55       &   24   & 6.31   &   K7     & 0.88   & 0.43       &  0.174 &     --    &   --   &  22.8        & CH10, SONG  \\
100 & HIP  92919                 & 18:55:53.23 &  +23:33:23.93   &  21.40       &  500       &  5.76   &  K0      & 0.80 & 0.37       &   --       & 2.88     & 0.00      &  0.10      & J13  \\ 
101 & HIP 94050              & 19:08:50.45 &  -42:25:41.50   &  33.84      &  400   &  5.97  & K1.5V     & 0.94 & $\sim$0.7  &   --       &  --      & --        &  $<1$      & L05   \\
102 & HIP  94863                 & 19:18:12.64 &  -38:23:04.45   &  41.90       & 4000       & 6.84   & G8V+K5V  & 0.83  & 0.63       &   --       & 2.50     & 0.00     &   0.09      & L05   \\
103 & HIP  95149                 & 19:21:29.80 &  -34:59:00.50   &  18.83       &  320   & 5.00  & G1V       & 0.89 & $<0.68$ &   --       &   --     &  --      &  $\sim20$   & B07   \\
104 & HIP 97255            & 19:45:57.35 &  +04 14 54.56   & 31.00        & 125        & 5.62  & G0V       & 1.10     & --     & --       & 3990      & --         &  21.1       & B07 \\  %%% NEW
105 & 2MASS J19560294-3207186     & 19:56:02.938 & -32:07:18.73   & 55.0         &  24        &   8.34    & M4        &    0.20      &  0.10      &  0.20         &  --         &    --        &       42.0      & L15 \\ %%%%NEW 
106 & HIP  100751                & 20:25:38.90 &  -56:44:06.00   &  54.82       &   30   & 2.46  & B7        & 5.82  &  0.26      &   --       &  11.7    & 0.0      &   0.32       & CH10  \\ 
107 & HIP  101800                & 20:37:49.12 &  +11:22:39.64   &  57.90       &  320       & 5.37    & A2V     & 2.00  &  0.49      &   --       & 11.0     & 0.23     &   0.40      & R13, N13  \\
108 & TYC 5206-0915-1            & 21:18:33.50 &  -06:31:44.00   &  76.40       &  250       & 8.17  & K1IV      & 0.91  &  0.45      &   --       &  --      &  --      &  $<10$     & NLP \\
109 & HIP  105404                & 21:20:59.80 &  -52:28:40.10   &  45.15       &   45       & 6.70  & K0V       & 0.90 & 0.80   &   -- &  1223     & 0.60     &  10.23     & CH10, SONG  \\
110 & HIP  105441            & 21:21:24.49 &  -66:54:57.37   &  30.17       &  4000  & 6.50 & K2.5Vk     & 0.85  & 0.42      &   --       &   --     &  --      &   $<10$     & K07   \\
111 & HIP  107556                & 21:47:02.44 &  -16:07:38.23   &  11.87       &  540       & 2.01  & A5        &  1.50 & 0.56       &  --        &  1.0     &  0.01    &  0.06      & N13   \\
112 & FS 1136                    & 21:49:06.20 &  -64:12:55.00   &  25.00   &  100       & 9.80  & M5        & 0.2   & 0.2        &  0.08      &   --     &  --      &   7.3     & CH10  \\% updated
113 & HIP 108195             & 21:55:11.40 &  -61:53:12.00   & 46.47    &   45       & 5.23   &    F1III       & 1.5    & 1.50       &  0.273     &  10070   & 0.546   &  46.8     & CH10  \\
114 & HIP  109901                & 22:15:35.20 &  -39:00 51.00   &  56.10       &  100       & 7.120  & K0V      &  0.89 & 0.45       &  --        &   --     &  --      &  $<10$     & SONG \\
115 & GJ 860                 & 22:27:59.47 &   57:41:45.15   &   4.00       &    1000           &  5.04 &   M2+M4          &   0.27     &   0.18     &   2.41    &    44.6yr    & 0.41  &   34.5   & H10 \\  
116 & PPM 366328             & 23:15:01.14 &  -63:34:24.54   &  60.00   & 4000    & 7.17 & K0         & 0.94   & 0.88   &  --   &   --     &  --      &    $<1$    & K07   \\ 
117 & HIP 116003                 & 23:30:13.4  &  -20:23:27.1    &   15.2       &   42        &   6.61    & M3        &  0.30    &      0.20    &  --   &   --     &  --      &   $<$5       & L15 \\  %%% new justine

\end{longtable} 	

\end{landscape} 
}}